\documentclass[twocolumn,preprintnumbers,superscriptaddress,nofootinbib,aps,prd,floatfix]{revtex4}

\usepackage{footmisc,enumerate}
\usepackage{subfigure}
\usepackage{amsmath,amssymb}
\usepackage{graphicx}
\usepackage{bbm,slashed}
\usepackage{xspace,slashed}
\usepackage{hyperref}
\usepackage{multirow}
\hypersetup{colorlinks=true, citecolor=blue, urlcolor=blue, linkcolor=blue}
\usepackage[normalem]{ulem}
\usepackage{tikz}

\newcommand{\OphiRD}{\mathcal{O}_{\phi \mathcal{R} \mathcal{D}}}
\newcommand{\CphiRD}{\mathcal{C}_{\phi \mathcal{R} \mathcal{D}}}

\newcommand{\CphiRDbar}{\bar{\mathcal{C}}_{\phi \mathcal{R} \mathcal{D}}}
\newcommand{\CphiSDbar}{\bar{\mathcal{C}}_{\phi \mathcal{S} \mathcal{D}}}
\newcommand{\massD}{M_{r^{\pm \pm}}}

\newcommand{\fR}{f_{\mathcal{R}}}
\newcommand{\fS}{f_{\mathcal{S}}}
\newcommand{\Mbar}{\overline{M}}
\newcommand{\Cerphi}{\mathcal{C}_{e \mathcal{R}\phi}}
\newcommand{\Clrphi}{\mathcal{C}_{\ell \phi \mathcal{R}}}
\newcommand{\Crle}{  \mathcal{C}_{\mathcal{R}\ell e}}

\newcommand{\MS}{$\overline{\text{MS}}$}
\newcommand{\+}{\pm}

\begin{document}
	
\newcommand{\og}{\ensuremath{\tilde{O}_g}\xspace}
\newcommand{\ot}{\ensuremath{\tilde{O}_t}\xspace}
\def\ani#1{\textcolor{red}{#1}}
\providecommand{\abs}[1]{\lvert#1\rvert}
\newcommand{\Znunujets}{(Z\to{\nu\bar{\nu}})+\text{jets}}
\newcommand{\Welnujets}{(W\to{\ell\nu})+\text{jets}}
\newcommand{\Znunujet}{(Z\to{\nu\bar{\nu}})+\text{jet}}
\newcommand{\Welnujet}{(W\to{\ell\nu})+\text{jet}} 

\newcommand{\cw}{\ensuremath{C_{\widetilde{W}}\xspace}}
\newcommand{\chwb}{\ensuremath{C_{H\widetilde{W}B}}\xspace}

\title{Effective connections of $a_\mu$, Higgs physics, and the collider frontier}

\begin{abstract}
We consider scalar extensions of the SM and their effective field theoretic generalisations to illustrate the phenomenological
connection between precision measurements of the anomalous magnetic
moment of the muon $a_\mu$, precision Higgs measurements, and direct collider sensitivity. To this end,
we consider charged BSM scalar sectors of the Zee-Babu type for which we develop a consistent, and 
complete dimensions-5 and -6 effective field theory extension. This enables us to track generic new physics effects that interact with the SM predominantly via radiative interactions. While the operator space is high dimensional, the intersection of exotics searches at the Large Hadron Collider, Higgs signal strength and anomalous muon magnetic measurements is manageably small. We find that consistency of LHC Higgs observations and $a_\mu$ requires a significant deformation of the new states' electroweak properties. Evidence in searches for doubly charged scalars as currently pursued by the LHC experiments can be used to further tension the BSMEFT parameter space and resolve blind directions in the EFT-extended Zee-Babu scenario.
\end{abstract}

\author{Anisha} \email{anisha@iitk.ac.in}
\affiliation{Indian Institute of Technology Kanpur, Kalyanpur, Kanpur 208016, India\\[0.1cm]}
\author{Upalaparna~Banerjee} \email{upalab@iitk.ac.in}
\affiliation{Indian Institute of Technology Kanpur, Kalyanpur, Kanpur 208016, India\\[0.1cm]}
\author{Joydeep~Chakrabortty} \email{joydeep@iitk.ac.in}
\affiliation{Indian Institute of Technology Kanpur, Kalyanpur, Kanpur 208016, India\\[0.1cm]}
\author{Christoph~Englert} \email{christoph.englert@glasgow.ac.uk}
\affiliation{School of Physics \& Astronomy, University of Glasgow, Glasgow G12 8QQ, United Kingdom\\[0.1cm]}
\author{Michael~Spannowsky} \email{michael.spannowsky@durham.ac.uk}
\affiliation{Institute for Particle Physics Phenomenology, Durham University, Durham DH1 3LE, United Kingdom\\[0.1cm]}
\author{Panagiotis~Stylianou}\email{p.stylianou.1@research.gla.ac.uk} 
\affiliation{School of Physics \& Astronomy, University of Glasgow, Glasgow G12 8QQ, United Kingdom\\[0.1cm]}

\preprint{IPPP/21/19}

\maketitle
\section{Introduction}
The search for new physics beyond the Standard Model (SM), albeit so far unsuccessful at the Large Hadron Collider (LHC), is key to the current particle physics phenomenology programme. The recent measurement of the anomalous muon magnetic moment 
\begin{equation}
a_\mu={(g-2)_\mu\over 2}\,,
\end{equation}
at Fermilab \cite{Muong-2:2021ojo} aligns with the previous results obtained at the BNL E821 experiment~\cite{Muong-2:2004fok}, leading to a  $\sim 4\sigma$ discrepancy~\cite{Davier:2019can,Aoyama:2020ynm} 
\begin{equation}\label{eq:1}
\Delta a_\mu = a_\mu(\text{exp}) - a_\mu(\text{SM}) = (25.1 \pm 5.9) \times 10^{-10}\,.
\end{equation}
While this deviation is a long standing, and potentially tantalising hint for the existence of new interactions beyond the SM that deserves further scrutiny from all angles~(see e.g.~\cite{Cowan:2021sdy}), it is flanked by broad consistency of collider measurements with the SM. In particular, this includes an increasing statistical control in searches for new heavy BSM states, and an enhanced precision in BSM tell-tale modifications of, e.g., precision Higgs data. 

On the one hand, one interpretation of this result is a large scale separation between the SM and BSM interactions, perhaps in the range $\Lambda\gtrsim 10~\text{TeV}$~\cite{Baer:2021aax,Athron:2021iuf,Frank:2021nkq,Ellis:2021zmg,Zhang:2021dgl,Jueid:2021avn,Altmannshofer:2021hfu,Chakraborti:2021squ,Chakraborti:2021dli}. On the other hand, we could be looking at an intricate cancellation between new physics effects that manifest themselves in the phenomenological outcome that we currently observe. 

In this paper, we elaborate on the latter option by performing a case study of the interplay of Higgs precision physics, $a_\mu$ and direct LHC sensitivity for a scenario that turns out to be particularly motivated for this purpose: For the Zee-Babu model~\cite{Zee:1985id,Babu:1988ki,Okada:2014qsa}, when extended by effective interactions, the phenomenological overlap of these three searches is particularly transparent. This enables us to discuss implications of low-energy precision measurements for high-energy observations, connecting anomalies around at the muon mass scale to TeV scale Higgs physics and the high energy exotics searches.

Many BSM theories contain charged scalar states. If we observe only these charged particles at the LHC when the rest of the spectrum is too heavy to be produced on-shell, then we can extend the Standard Model Effective Field Theory (SMEFT) with these additional TeV scale degrees of freedom, leading to a BSMEFT scenario~\cite{Banerjee:2020jun}. If such a case is realised in nature, constraints of the BSMEFT Wilson coefficients are required to gain a qualitative understanding of the new physics energy scales that perhaps lie beyond the reach of the LHC.

After electroweak symmetry breaking, the Zee-Babu model is the simplest framework that provides singly and a doubly charged scalars that also addresses open questions in neutrino physics. We choose this spectrum to construct a prototype BSMEFT scenario, which, as we show, is particularly suited to discuss the phenomenological interplay of Higgs physics, LHC exotics searches and anomalous magnetic moment studies. While we focus on the doubly charged state as a LHC smoking gun signature, the inclusion of the singly charged scalar is crucial to the comparison of $a_\mu$, Higgs data and future direct sensitivity.

\begin{figure*}[!t]
	\parbox{0.3\textwidth}{
		\vspace{3cm}
		\caption{BSM Feynman diagrams contributing to the muon anomalous magnetic moment $\mu\to \mu\gamma$ via the new propagating $r^{\pm\pm}$ and its EFT interactions. The vertices include the renormalisable and the dimension-6 interactions. Similar diagrams arise from the $h^\+$ scalar.
			\label{fig:feyn_amu_rpp}}}
	\hskip 0.03\textwidth
	\parbox{0.65\textwidth}{\includegraphics[width=0.65\textwidth]{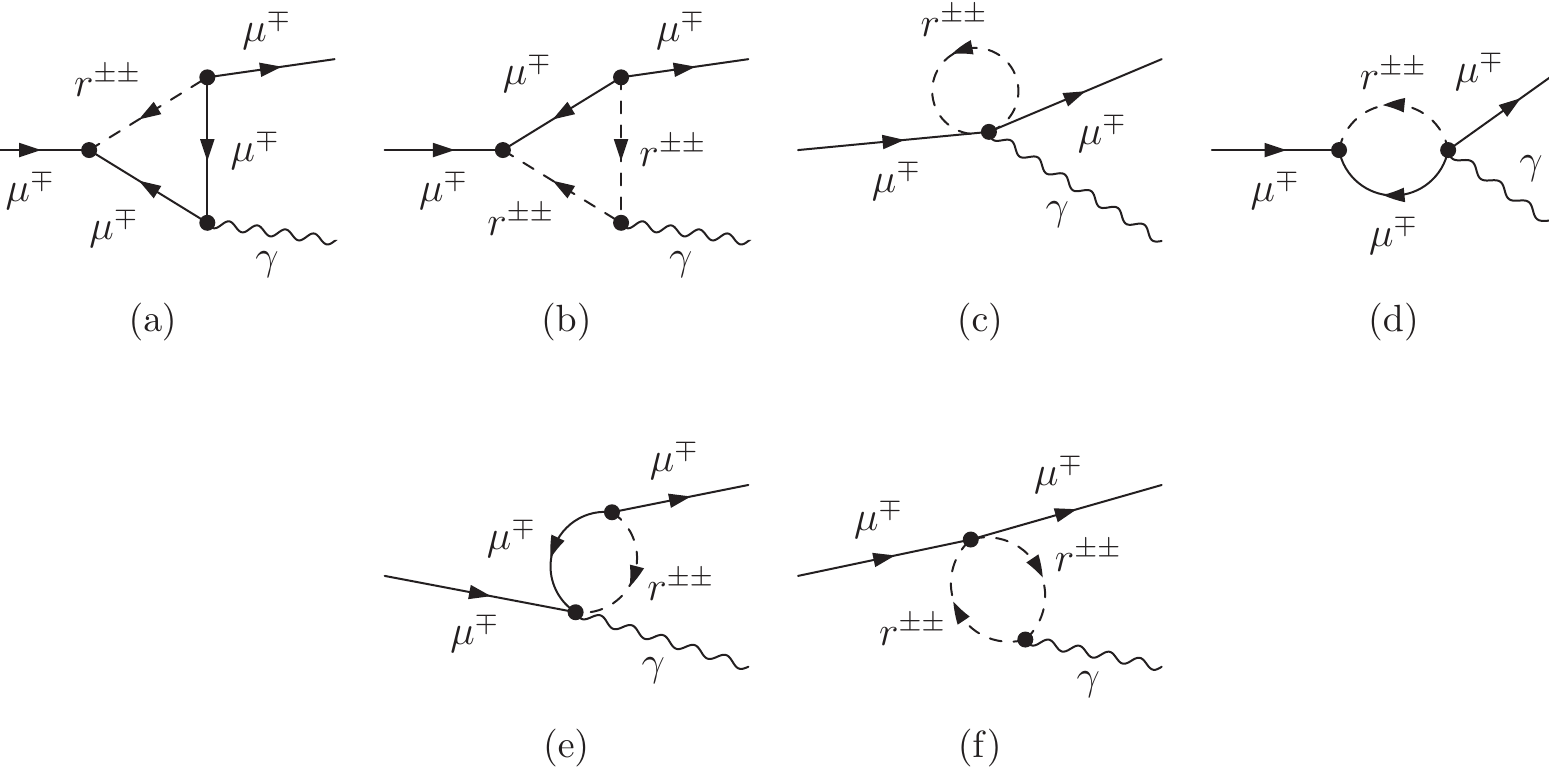}}
\end{figure*}

This work is organised as follows: In Sec.~\ref{sec:model}, we briefly review the Zee-Babu model before providing a detailed discussion of its dimension-5 and -6 effective field theory (EFT) extension. In Sec.~\ref{sec:pheno}, we turn to the phenomenological implications that we focus on in this paper, i.e. the anomalous muon magnetic moment in Sec.~\ref{sec:gm2}, expected modifications of 125 GeV Higgs boson measurements in Sec.~\ref{sec:hdec}, and the direct sensitivity to doubly charged scalar bosons as a smoking gun of this scenario in Sec.~\ref{sec:hppdec}. In Sec.~\ref{sec:bsmeftint}, we combined these three searches to highlight their complementarity and intersection. We conclude in Sec.~\ref{sec:conc}.
	
\section{The model}
\label{sec:model}
The Zee-Babu model~\cite{Zee:1985id,Babu:1988ki} is an extension of the usual SM Lagrangian by two $SU(2)_L$ and color singlet scalar fields with non-trivial hypercharges
\begin{equation}
\label{eq:2}
\begin{split}
\mathcal{S} &: (\textbf{1},\textbf{1},1)\,,\\
\mathcal{R} &: (\textbf{1},\textbf{1},2)\,.
\end{split}
\end{equation}
These give rise to the new renormalisable and effective interactions determined by the gauge symmetry $SU(3)_C\otimes SU(2)_L\otimes U(1)_Y$.
The renormalisable Lagrangian is given by
\begin{eqnarray}\label{eq:3}
	\mathcal{L}_{\text{renorm}} &=& - \frac{1}{4}G^{A}_{\mu\nu}G^{A\mu\nu} - \frac{1}{4}W^{I}_{\mu\nu}W^{I\mu\nu} - \frac{1}{4}B_{\mu\nu}B^{\mu\nu}\nonumber\\ 
	& &\hspace{-6mm}+(\mathcal{D}_{\mu} \phi)^{\dagger}(\mathcal{D}^{\mu} \phi)+ (\mathcal{D}_{\mu} \mathcal{S})^{\dagger}(\mathcal{D}^{\mu}\, \mathcal{S})+ (\mathcal{D}_{\mu} \mathcal{R})^{\dagger}(\mathcal{D}^{\mu}\, \mathcal{R})\nonumber\\
	& &\hspace{-6mm}-\mathcal{V}(\phi,\mathcal{S},\mathcal{R})
	+i(\overline{L}\gamma^{\mu}\mathcal{D}_{\mu}L+\overline{e}\gamma^{\mu}\mathcal{D}_{\mu}e+\overline{Q}\gamma^{\mu}\mathcal{D}_{\mu}Q\nonumber\\
	& &\hspace{-6mm}+\overline{u}\gamma^{\mu}\mathcal{D}_{\mu}u+\overline{d}\gamma^{\mu}\mathcal{D}_{\mu}d)
	+ (\mathcal{L}_{\text{Yukawa}}+\text{h.c.})\, ,
\end{eqnarray}
where 
\begin{eqnarray}\label{eq:4}
	G^{A}_{\mu\nu} &=& \partial_{\mu} G^{A}_{\nu} - \partial_{\nu} G^{A}_{\mu} + g_3 f^{ABC} G^{B}_{\mu}G^{C}_{\nu},\nonumber\\
	W^{I}_{\mu\nu} &=& \partial_{\mu} W^{I}_{\nu} - \partial_{\nu} W^{I}_{\mu} + g \epsilon^{IJK} W^{J}_{\mu}W^{K}_{\nu},\nonumber\\
	B_{\mu\nu} &=& \partial_{\mu} B_{\nu} - \partial_{\nu} B_{\mu},
\end{eqnarray}
are the field strength tensors corresponding to $SU(3)_{C}$, $SU(2)_{L}$, and $U(1)_{Y}$ respectively, here $\{A, B, C\} \in \{1,2,\cdots,8\}$, and $\{I,J,K\} \in \{1,2,3\}$.	
	
The scalar potential $\mathcal{V}(\phi,\mathcal{S},\mathcal{R})$ in Eq.~\eqref{eq:2} reads
\begin{eqnarray}\label{eq:5}
	\mathcal{V}(\phi,\mathcal{S},\mathcal{R}) &=& \mu_{1}^{2} \,(\phi^{\dagger}\phi)+ \mu_{2}^{2} \,(\mathcal{S}^{\dagger}\mathcal{S})+ \mu_{3}^{2} \,(\mathcal{R}^{\dagger}\mathcal{R})\nonumber\\
	& &+m\,(\mathcal{S}^2 \,\mathcal{R}^{\dagger}+\mathcal{(S^{\dagger})}^2 \,\mathcal{R})+\lambda_{1}(\phi^{\dagger}\phi)^{2}\nonumber\\
	& &+\,\lambda_{2}\,(\mathcal{S}^{\dagger}\mathcal{S})^{2}+\,\lambda_{3}\,(\mathcal{R}^{\dagger}\mathcal{R})^{2}\nonumber\\
	& &+\lambda_{4}\,(\phi^{\dagger}\phi)(\mathcal{S}^{\dagger}\mathcal{S})+\lambda_{5}\,(\phi^{\dagger}\phi)(\mathcal{R}^{\dagger}\mathcal{R})\nonumber\\
	& &+\lambda_{6}\,(\mathcal{S}^{\dagger}\mathcal{S})(\mathcal{R}^{\dagger}\mathcal{R})\,.
\end{eqnarray}
In contrast to the SM, the quantum numbers of the Zee-Babu singlet scalars allow new quartic as well as trilinear interactions, as can be seen in Eq~\eqref{eq:5}. We have denoted the new quartic couplings as $\lambda_{2}$, $\lambda_{3}$, $\lambda_{4}$, $\lambda_{5}$, and $\lambda_{6}$, whereas $m$ parametrises the trilinear scalar interaction.   
	
$\mathcal{L}_{\text{Yukawa}}$ contains two new Yukawa-like interactions along with the usual SM ones
\begin{eqnarray}\label{eq:6}
\mathcal{L}_{\text{Yukawa}} &=& -y_{e}\overline{L}e\phi-y_{u}\overline{Q}u\tilde{\phi}-y_{d}\overline{Q}d\phi\nonumber\\
& &-f_{\mathcal{S}}(\overline{L^{c}}i\tau_{2}L)\mathcal{S}-f_{\mathcal{R}}(\overline{e^{c}}\,e)\mathcal{R}\,.
\end{eqnarray} 
Here, $\tilde{\phi}_{i}=\epsilon_{ij} \phi_{j}^{\ast}$ is the charge-conjugated Higgs doublet. The Yukawa couplings $f_{\mathcal{S}}$ and $f_{\mathcal{R}}$ parametrise the interactions between $SU(2)_{L}$ lepton doublet $L$, singlet $e$ with scalars $\mathcal{S}$,  and $\mathcal{R}$, respectively.

Bounds on the parameters in Eq.~\eqref{eq:5} can be derived from examining the shape of $\mathcal{V}(\phi,\mathcal{S},\mathcal{R})$. For the potential to be bounded from below each of $\lambda_{1},\lambda_{2}$ and $\lambda_{3}$ should be positive. To achieve the overall positivity of the potential one can find the following relations, see e.g.~\cite{Herrero-Garcia:2014hfa}
\begin{multline}\label{eq:7}
		\lambda_{4}/2\sqrt{\lambda_{1}\lambda_{2}} > -1\,, \quad \lambda_{5}/2\sqrt{\lambda_{1}\lambda_{3}} > -1\,,\\
		\lambda_{6}/2\sqrt{\lambda_{2}\lambda_{3}} > -1\,.
\end{multline}
Apart from that, there are $\sim 4\pi$ perturbativity bounds for $\lambda_{i}$, ${i=2,\dots,6}$.
	
After electroweak symmetry breaking, $\phi$ acquires a vacuum expectation value (vev) and gives rise to the physical Higgs $H$. $\mathcal{S}$ and $\mathcal{R}$ emerge as singly and doubly charged scalars $h^{\pm}$ and $r^{\pm\pm}$, respectively.
\medskip
	
We aim to track the generic physics that predominantly couples to $\mathcal{S},\mathcal{R}$. To this end, we modify SM correlations not only through the presence of $h^{\pm}$ and $r^{\pm\pm}$, but also include the interactions that arise from integrating out the new physics that further deform the $\mathcal{S},\mathcal{R}$ interactions with SM matter. We therefore extend the renormalisable Lagrangian with a complete, independent, and exhaustive set of dimensions-5 and -6 effective operators
\begin{eqnarray}\label{eq:8}
	\mathcal{L} = \mathcal{L}_{\text{renorm}} +
	\sum_{j=1}^{N}\frac{\mathcal{C}^{(5)}_j}{\Lambda}\mathcal{O}^{(5)}_j + \sum_{k=1}^{M}\frac{\mathcal{C}^{(6)}_k}{\Lambda^{2}}\mathcal{O}^{(6)}_k\,.
\end{eqnarray}
	
We choose to express the operator sets using the Warsaw basis methodology~\cite{Grzadkowski:2010es}. The complete set of effective operators that couple $\mathcal{S}$ and $\mathcal{R}$ to the SM fields have been listed in appendix~\ref{sec:App_A}. For the purpose of our work, we consider only those operators that affect the anomalous magnetic moment for muon, the loop-induced neutral Higgs decays, and production and decay of the charged scalars. The gauge invariant structures for these operators are given in Tab.~\ref{table:ops_structures}. Throughout this paper, we will consider real values for the Wilson coefficients ($\mathcal{C}_{_k}$) alongside a trivial flavour structure of the new interactions.\footnote{$f_{\mathcal{R}}$ and $f_{\mathcal{S}}$ symmetric and anti-symmetric couplings in lepton flavour space which project out these related combinations of Wilson coefficients in concrete calculations.}
	
\begin{table}[!t]
	\centering
	\renewcommand{\arraystretch}{1.9}
	{\scriptsize\begin{tabular}{||c|c||c|c||}
			\hline
			\hline
			\multicolumn{2}{||c||}{$\Phi^5$}&
			\multicolumn{2}{c||}{$\Psi^2\Phi^2$}
			\\
			\hline
			$\mathcal{O}_{r}$&$\boldsymbol{(\phi^{\dagger}\,\phi)\,\mathcal{R}^{\dagger}\,\mathcal{S}^2}$&$\mathcal{O}_{le\phi \mathcal{S}}$&$\boldsymbol{\overline{L}\,e\,\widetilde{\phi}\mathcal{S}}$\\
			\hline
			\multicolumn{2}{||c||}{$\Phi^4\mathcal{D}^2$}&
			\multicolumn{2}{c||}{$\Phi^6$}
			\\
			\hline
			$\mathcal{O}_{\phi\mathcal{R}\mathcal{D}}$ & $(\phi^{\dagger}\,\phi)\,\left[(\mathcal{D}^{\mu}\,\mathcal{R})^{\dagger}(\mathcal{D}_{\mu}\,\mathcal{R})\right]$& $\mathcal{O}_{\phi\mathcal{R}}$ & $(\phi^{\dagger} \,\phi)^2 \,(\mathcal{R}^{\dagger} \,\mathcal{R})$\\
			$\mathcal{O}_{\phi\mathcal{S}\mathcal{D}}$ & $(\phi^{\dagger}\,\phi)\,\left[(\mathcal{D}^{\mu}\,\mathcal{S})^{\dagger}(\mathcal{D}_{\mu}\,\mathcal{S})\right]$&$\mathcal{O}_{\phi\mathcal{S}}$ & $(\phi^{\dagger} \,\phi)^2 \,(\mathcal{S}^{\dagger} \,\mathcal{S})$\\
			$\mathcal{O}_{\mathcal{R}\phi\mathcal{D}}$ & $(\mathcal{R}^{\dagger}\,\mathcal{R})\,\left[(\mathcal{D}^{\mu}\,\phi)^{\dagger}(\mathcal{D}_{\mu}\,\phi)\right]$& &\\
			$\mathcal{O}_{\mathcal{S}\phi\mathcal{D}}$ & $(\mathcal{S}^{\dagger}\,\mathcal{S})\,\left[(\mathcal{D}^{\mu}\,\phi)^{\dagger}(\mathcal{D}_{\mu}\,\phi)\right]$ & &\\
			
			\hline
			\multicolumn{2}{||c||}{$\Psi^2\Phi^2\mathcal{D}$}&
			\multicolumn{2}{c||}{$\Psi^2\Phi^3$}
			\\
			\hline
			$\mathcal{O}_{\mathcal{R}l e}$ & $\boldsymbol{(\overline{L^c}\,\gamma^{\mu}\,e)(\phi\,i\mathcal{D}_{\mu}\,\mathcal{R})}$&$\mathcal{O}_{ l\phi \mathcal{S}}$ & $\boldsymbol{(\overline{L^{c}}i\tau_{2}L)\,(\phi^{\dagger}\,\phi)\,\mathcal{S}}$\\
			$\mathcal{O}_{\mathcal{S}l e}$ & $\boldsymbol{(\overline{L^c}\,\gamma^{\mu}\,e)(\widetilde{\phi}\,i\mathcal{D}_{\mu}\,\mathcal{S})}$&$\mathcal{O}_{l \phi \mathcal{R}}$ & $\boldsymbol{(\overline{L^{c}}i\tau_{2}L)\,(\phi^{\dagger}\,\mathcal{R}\,\widetilde{\phi})}$\\
			$\mathcal{O}_{\mathcal{R}q}$ & $(\overline{Q} \,\gamma^{\mu} \,Q) \,(\mathcal{R}^{\dagger} i\overleftrightarrow{\mathcal{D}}_{\mu}\mathcal{R})$ &$\mathcal{O}_{e \mathcal{R} \phi}$ &$\boldsymbol{(\phi^{\dagger}\phi)\,\mathcal{R}\,(\overline{e^c}\,e)}$\\
			$\mathcal{O}_{\mathcal{S}q}$ & $(\overline{Q} \,\gamma^{\mu} \,Q) \,(\mathcal{S}^{\dagger} i\overleftrightarrow{\mathcal{D}}_{\mu}\mathcal{S})$ &$\mathcal{O}_{u \phi \mathcal{R}}$ & $\boldsymbol{(\overline{Q}\,u)\,\widetilde{\phi}\,(\mathcal{R}^{\dagger}\mathcal{R})}$  \\
			$\mathcal{O}_{\mathcal{R}u}$ & $(\overline{u} \,\gamma^{\mu} \,u) \,(\mathcal{R}^{\dagger} i\overleftrightarrow{\mathcal{D}}_{\mu}\mathcal{R})$ &$\mathcal{O}_{u \phi \mathcal{S}}$&$\boldsymbol{(\overline{Q}\,u)\,\widetilde{\phi}\,(\mathcal{S}^{\dagger}\mathcal{S})}$ \\
			$\mathcal{O}_{\mathcal{S}u}$ & $(\overline{u} \,\gamma^{\mu} \,u) \,(\mathcal{S}^{\dagger} i\overleftrightarrow{\mathcal{D}}_{\mu}\mathcal{S})$& & \\
			\hline
			\multicolumn{4}{||c||}{$\Phi^2X^2$}\\
			\hline
			$\mathcal{O}_{ B \mathcal{R}}$ & $B_{\mu\nu} \,B^{\mu\nu} \,(\mathcal{R}^{\dagger} \,\mathcal{R})$ &$\mathcal{O}_{ \widetilde{B} \mathcal{R}}$ & $\widetilde{B}_{\mu\nu} \,B^{\mu\nu} \,(\mathcal{R}^{\dagger} \,\mathcal{R})$ \\
			$\mathcal{O}_{ B \mathcal{S}}$ &
			$B_{\mu\nu} \,B^{\mu\nu} \,(\mathcal{S}^{\dagger} \,\mathcal{S})$& $\mathcal{O}_{ \widetilde{B} \mathcal{S}}$ & $\widetilde{B}_{\mu\nu} \,B^{\mu\nu} \,(\mathcal{S}^{\dagger} \,\mathcal{S})$ \\
			$\mathcal{O}_{ W \mathcal{R}}$ &
			$W^{I}_{\mu\nu} \,W^{I\mu\nu} \,(\mathcal{R}^{\dagger} \,\mathcal{R})$&$\mathcal{O}_{ \widetilde{W} \mathcal{R} }$&
			$\widetilde{W}^{I}_{\mu\nu} \,W^{I\mu\nu} \,(\mathcal{R}^{\dagger} \,\mathcal{R})$ \\
			$\mathcal{O}_{ W \mathcal{S}}$ &
			$W^{I}_{\mu\nu} \,W^{I\mu\nu} \,(\mathcal{S}^{\dagger} \,\mathcal{S})$ &$\mathcal{O}_{ \widetilde{W} \mathcal{S} }$&
			$\widetilde{W}^{I}_{\mu\nu} \,W^{I\mu\nu} \,(\mathcal{S}^{\dagger} \,\mathcal{S})$\\
			$\mathcal{O}_{G\mathcal{R}}$ & $G^{A}_{\mu\nu}\,G^{A\mu\nu}\,(\mathcal{R}^{\dagger} \,\mathcal{R})$ & $\mathcal{O}_{\widetilde{G}\mathcal{R}}$ & $\widetilde{G}^{A}_{\mu\nu}\,G^{A\mu\nu}\,(\mathcal{R}^{\dagger} \,\mathcal{R})$\\
			$\mathcal{O}_{G\mathcal{S}}$ & $G^{A}_{\mu\nu}\,G^{A\mu\nu}\,(\mathcal{S}^{\dagger} \,\mathcal{S})$ & $\mathcal{O}_{\widetilde{G}\mathcal{S}}$ & $\widetilde{G}^{A}_{\mu\nu}\,G^{A\mu\nu}\,(\mathcal{S}^{\dagger} \,\mathcal{S})$\\
			\hline
			\multicolumn{4}{||c||}{$\Psi^2\Phi X$}\\
			\hline
			
			$\mathcal{O}_{e B \mathcal{S}} $&
			$\boldsymbol{B_{\mu\nu} \,(\overline{L^{c}}\,\sigma^{\mu\nu} \,L) \,\mathcal{S}}$&
			$\mathcal{O}_{e W \mathcal{S}} $&
			$\boldsymbol{W^I_{\mu\nu} \,(\overline{L^{c}}\,\tau^I\,\sigma^{\mu\nu} \,L) \,\mathcal{S}}$\\
			\hline
	\end{tabular}}
	\caption{Explicit structures of the dimension-5 and -6 operators contributing to muon anomalous magnetic moment, loop-induced Higgs decay and production and decay for $h^{\pm}$ and $r^{\pm\pm}$. The operators in bold have distinct hermitian conjugates. $A \in \{1,2,\cdots,8\}$ and $I \in \{1,2,3\}$.}\label{table:ops_structures}
\end{table}

\section{Phenomenology}	
\label{sec:pheno}
\subsection{Muon anomalous magnetic moment}
\label{sec:gm2}
We first calculate the anomalous magnetic moment for muon for the considered scenario, extending well-documented results~\cite{Leveille:1977rc,Moore:1984eg,Nebot:2007bc,Schmidt:2014zoa} to effective interactions. In Tab.~\ref{table:mu_g2}, we have listed the parameters and the operators contributing to $a_\mu$.
\begin{table}[h]
	\centering
	\renewcommand{\arraystretch}{1.9}
	{\scriptsize\begin{tabular}{||c|c|c||}
			\hline
			\hline
			Charged scalar & Renormalisable & Contributing \\
			type & couplings & operators\\
			\hline
			\hline
			\multirow{3}{*}{$h^{\pm}$ } & \multirow{2}{*}{$f_{\mathcal{S}}$} & $\mathcal{O}_{\phi \mathcal{S}\mathcal{D}}$,\hspace{2mm}$\mathcal{O}_{ e W \mathcal{S}}$,\\
			& &$\mathcal{O}_{ e B \mathcal{S}}$,\hspace{2mm}$\mathcal{O}_{\mathcal{S}l e}$,\\
			& &$\mathcal{O}_{ l\phi \mathcal{S}}$.\\
			\hline
			\multirow{2}{*}{$r^{\pm\pm}$} & \multirow{2}{*}{$f_{\mathcal{R}}$} & $\mathcal{O}_{\phi \mathcal{R}\mathcal{D}}$,\hspace{2mm}$\mathcal{O}_{e \mathcal{R} \phi }$,\\
			& &$\mathcal{O}_{\mathcal{R}l e}$,\hspace{2mm}$\mathcal{O}_{l \phi \mathcal{R}}. $\\
			\hline
	\end{tabular}}
	\caption{The renormalisable couplings and the singly- and doubly-charged scalar related operators that contribute to the muon anomalous magnetic moment.}
	\label{table:mu_g2}
\end{table}
\begin{figure*}[!t]
	\subfigure[\label{fig:amufr}]{\includegraphics[width=0.32\textwidth]{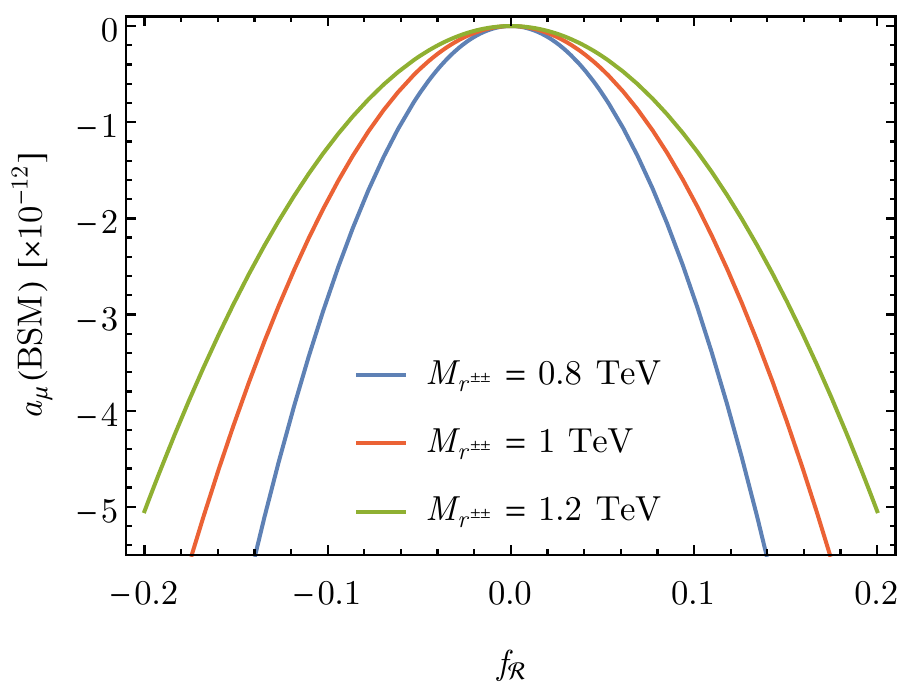}}\hfill
	\subfigure[\label{fig:amufs}]{\includegraphics[width=0.32\textwidth]{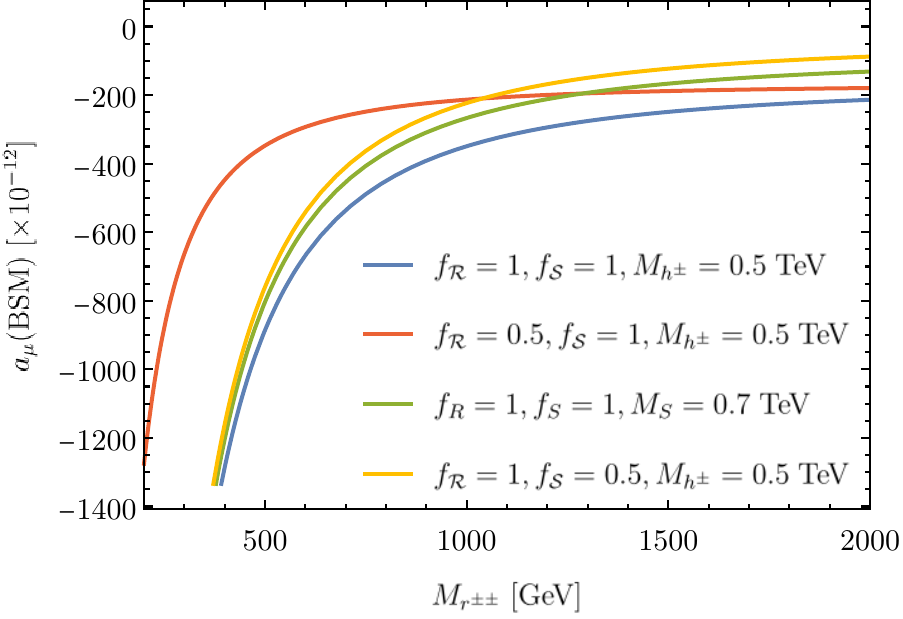}}\hfill
	\subfigure[\label{fig:amumr}]{\includegraphics[width=0.32\textwidth]{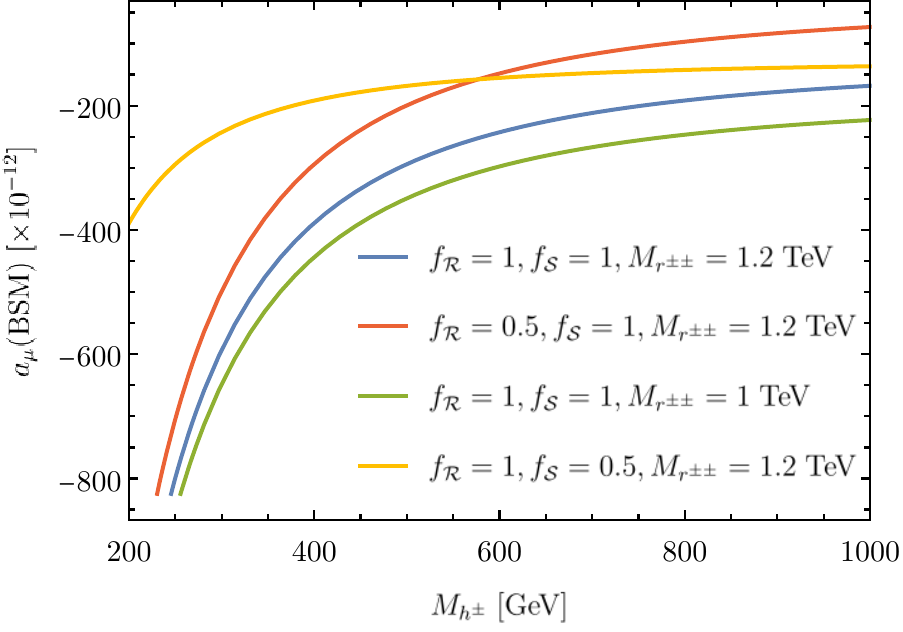}}\\
	\subfigure[\label{fig:amums}]{\includegraphics[width=0.32\textwidth]{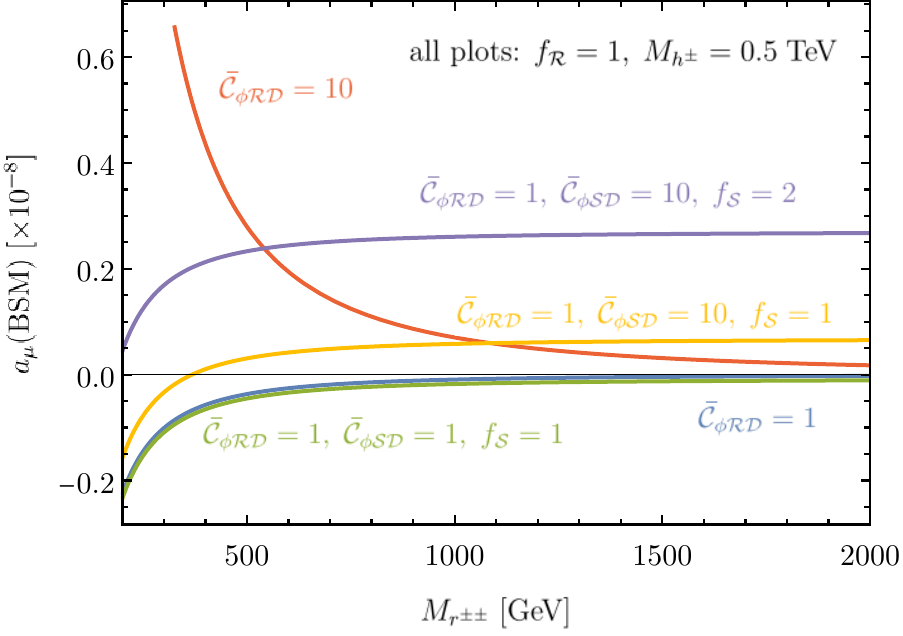}}\hfill
	\subfigure[\label{fig:amumrbsmeft}]{\includegraphics[width=0.32\textwidth]{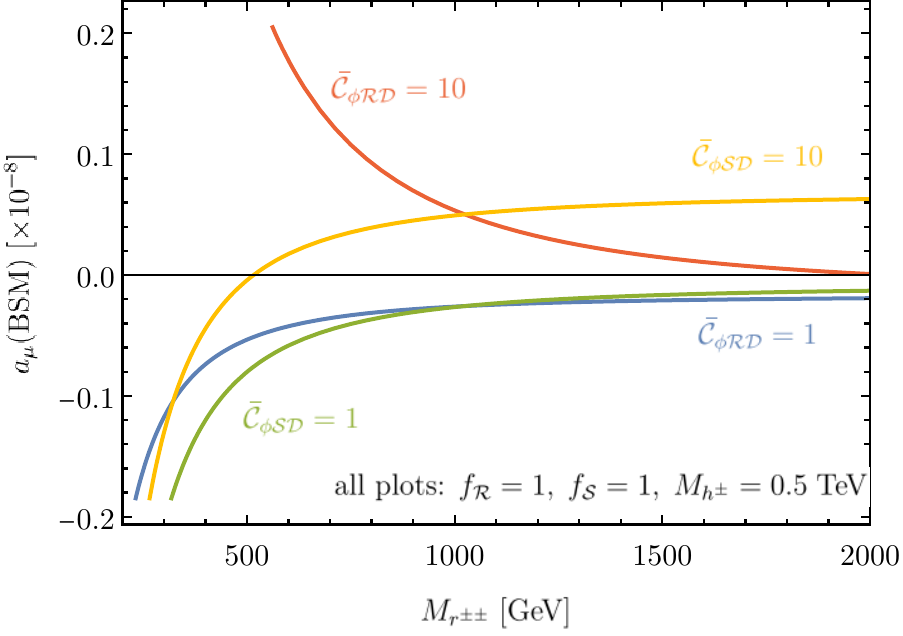}}\hfill
	\subfigure[\label{fig:amucphird}]{\includegraphics[width=0.32\textwidth]{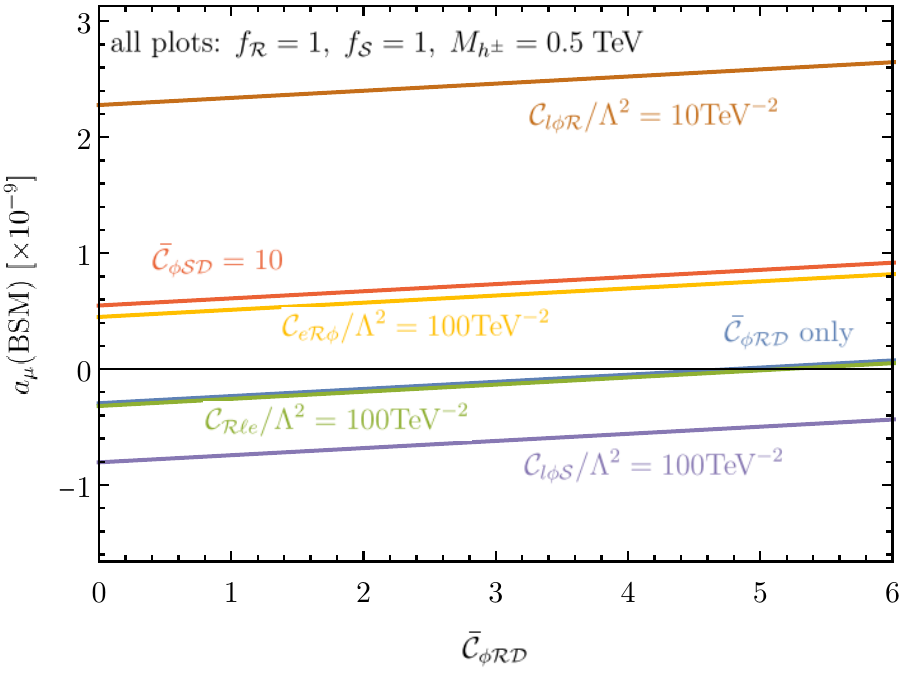}}\\
	\caption{Impact of various new physics parameters on $a_\mu$. The top row shows the dependence on terms from the renormalisable part of the Lagrangian~Eq.~\eqref{eq:3} while the bottom row includes effects from different effective operators. In (d) we investigate how large $\CphiRDbar$ is required to be in order to get a positive result when only the doubly charged state is present and we also show that $\CphiRDbar$ can be kept low by introducing the singly charged scalar. The effect of altering $C_{\phi \mathcal{R} \mathcal{D}}$ and $C_{\phi \mathcal{S} \mathcal{D}}$ is also shown in (e) when both scalars are included. Figure (f) shows the linear dependence of the anomalous magnetic moment on $C_{\phi \mathcal{R} \mathcal{D}}$ and how it is shifted when introducing additional operators (through we choose $\bar{\mathcal{C}}_i = \mathcal{C}_i v^2 / \Lambda^2$ for convenience).\label{fig:amu}}
\end{figure*}
	
The phenomenological appeal of $a_\mu$ is rooted in the fact that it provides an unambiguous BSM effect for UV-complete scenarios: When matching the contribution of a concrete BSM theory to the SMEFT operators that gives rise to $a_\mu$ (we refer the muon mass as $M_\mu$)
\begin{eqnarray}\label{eq:12}
	\Delta a_{\text{SMEFT}}	&=&  \,\frac{\sqrt{2} \,v \,M_{\mu}}{e} \mathcal{C}_{eA}\\
	&=& \frac{\sqrt{2} \,v \,M_{\mu}}{e} \big(\mathcal{C}_{eW}\,\sin{\theta_{_W}}\,-\,\mathcal{C}_{eB}\,\cos{\theta_{_W}}\big)\,,\nonumber
\end{eqnarray}
the Wilson coefficients $\mathcal{C}_{eW}$, $\mathcal{C}_{eB}$ of the operators  $\mathcal{O}_{eW}$ and $\mathcal{O}_{eB}$ in the language of Refs.~\cite{Grzadkowski:2010es,Brivio:2017btx,Dedes:2017zog}		
\begin{equation}
\label{eq:13}
\begin{split}
	\mathcal{O}_{eW} &= W^I_{\mu\nu} (\overline{L} \,\sigma^{\mu\nu} \,e) \,\tau^I \,\phi\,,\\
	\mathcal{O}_{eB} &= B_{\mu\nu} (\overline{L} \,\sigma^{\mu\nu} \,e)\,\phi\,,
\end{split}
\end{equation}
($ \sigma^{\mu\nu}= i[\gamma^\mu,\gamma^\nu]/2$) will remain finite to all orders in perturbation theory. For the EFT model discussed in the above section this remains true to one-loop order for a range of interactions, but broadly speaking, EFT insertions related to the SM or BSM particle content will generically imply a renormalisation of the operators related to $a_\mu$ as well. The precision of the obtained measurements of $a_\mu$ then motivates the inclusion of this observable to the defining input of the field theory to tension correlation predictions for other observables. In the following we will work in the mass-basis of the SM as indicated in Eq.~\eqref{eq:12}, where we consider $\mathcal{O}_{eA} = A_{\mu\nu} (\overline{e}\,\sigma^{\mu\nu}\,e)\,v$ (where $A_{\mu\nu}$ is the QED field strength).\footnote{Renormalisation of $Z-A$ mixing~\cite{Denner:1991kt,Denner:2019vbn} implies the requirement of considering the $Z$ boson-associated magnetic moment of the muon $\mathcal{O}_{eZ}$. In this work, however, we will focus on $a_\mu$, which means that the renormalisation procedure is confined to $\mathcal{C}_{eA}$ operator structures.}
	
Concretely, we evaluate the one-loop three-point vertex function 
\begin{equation}
\Gamma^\mu = -ie \bar u(p')\left[ \gamma^\mu F_1(k^2)   +  { i \over 2 M_\mu} \sigma^{\mu\nu} k_\nu F_2(k^2)+ \dots \right] u(p)\,,
\end{equation}
with momentum transfer $k=p'-p$. The ellipses denote additional form factors that appear in chiral gauge theories, e.g. the anomalous electric dipole moment. In this work we limit ourselves to the anomalous magnetic moment 
\begin{equation}
a_\mu=F_2(0)\,,
\end{equation} 
which is directly related to the effective Lagrangian of Eq.~\eqref{eq:12}. We employ dimensional regularisation and choose \MS~renormalisation for the Wilson coefficients and on-shell renormalisation for the remaining electroweak contributions, in particular for the external muon fields (see~\cite{Denner:1991kt} for a review); Feynman diagram contributions are shown in Fig.~\ref{fig:feyn_amu_rpp}. We consider terms up to $\sim 1/\Lambda^2$ (i.e. we truncate the series expansion at dimension-6 level), and renormalise the structure in Eq.~\eqref{eq:12} to cancel the divergence associated with the $\mathcal{C}_{eA}$ Lorentz structure (details are presented appendix~\ref{app:renorm}). At the considered one-loop, $\Lambda^{-2}$ level, these are exclusively given by the effective operator insertions related to $h^\pm$, the dimension-6 singularities of $a_\mu$ arise from $\sim {\cal{C}}_{e B \mathcal{S}},{\cal{C}}_{e W \mathcal{S}}$ loop contributions. We use {\sc{FeynArts}}~\cite{Hahn:2000kx} to enumerate the relevant one-loop diagrams and {\sc{FormCalc}}~\cite{Hahn:1998yk} for calculating the amplitudes and extracting the relevant form factor. {\sc{PackageX}}~\cite{Patel:2015tea} is used for simplifications of Passarino-Veltman scalar loop integrals~\cite{Passarino:1978jh}.

The anomalous magnetic moment in the context of the Zee-Babu model has been studied extensively in the past (see for example Refs.~\cite{Leveille:1977rc,Moore:1984eg,Nebot:2007bc,Schmidt:2014zoa}). We reproduce the standard result
\begin{equation}
	a^{\text{d4}}_\mu(\text{Zee-Babu}) = - \frac{M_\mu^2}{24 \pi^2} \left( \frac{(\fS^\dagger \fS)_{\mu\mu}}{M_{h^{\pm}}^2} + 4 \frac{ (\fR^\dagger \fR)_{\mu\mu}}{M_{r^{\pm\pm}}^2} \right) \,.
	\label{eq:zeebabu-a-mu}
\end{equation}
and the famous Schwinger result $\Delta a_\mu(\text{QED})=\alpha/2\pi$~\cite{Schwinger:1948iu} as a cross check and to align conventions.\footnote{} A summary of the impact of EFT operators, alongside the sensitivity to renormalisable couplings of the scenario introduced in Sec.~\ref{sec:model} is provided in Tab.~\ref{table:mu_g2}. The effect of different parameters on $a_\mu$ arising from the BSM contributions is shown in Fig.~\ref{fig:amu}.

The contributions from the renormalisable charged scalar interactions are negative, Eq.~\eqref{eq:zeebabu-a-mu}, which is also clearly visible from Figs.~\ref{fig:amu}(a)-(c). To explain the experimental measurement of the anomalous magnetic moment, which favours a positive $\Delta a_\mu(\text{BSM})$ slightly larger than the SM expectation, this negative contribution needs to be overcome by additional EFT contributions. These can be logarithmically enhanced for large mass gaps $M_{r^{\+\+}},M_{h^{\+}}\gg M_\mu$. The $a_\mu$ contributions for the effective interactions related to $r^{\pm\pm}$ take a particularly compact form in the limit $M_{r^{\+\+}}\gg M_\mu$
\begin{widetext}
\begin{multline}
\label{eq:d6amu}
 \Lambda^2 \times a^{\text{d6},r^{\+\+}}_\mu(\text{Zee-Babu})=	\frac{\fR M_{\mu}^2 v^2 (\Cerphi)_{\mu\mu}}{6 \pi ^2 \massD^2}+\frac{\fR M_{\mu}^2 v^2 (\Clrphi)_{\mu\mu}}{2 \pi ^2 \massD^2}\left(\log \left(\frac{\massD}{M_{\mu}}\right)-\frac{1}{4}\right)\\
+\frac{\fR M_{\mu}^3 v (\Crle)_{\mu\mu}}{\sqrt{2} \pi ^2  \massD^2} \left(\frac{7}{12} -\log \left(\frac{\massD}{M_{\mu}}\right)\right)+\frac{\fR^2 M_{\mu}^2 v^2 \CphiRD}{12 \pi ^2 \massD^2}\,.
\end{multline}
\end{widetext}
This together with the fully-renormalised $h^\+$ interactions give rise to the behaviour shown Figs.~\ref{fig:amu}(d)-(f).\footnote{It is worth highlighting that these distributions include the Yukawa interactions of $r^{\+\+},h^{\+}$, which means that there are non-vanishing BSM contributions to $a_\mu$ in all displayed cases.}

Attributing the observed $a_\mu$ to dominant $C_{\phi \mathcal{R} \mathcal{D}}$ interactions requires large Wilson coefficients, and we will discuss the phenomenological implication of such a scenario below. In parallel, when we consider the contributions related to $h^\+$, the observed $C_{\phi \mathcal{R} \mathcal{D}}-a_\mu$ correlation can be altered which again leads to experimentally testable implications (see Sec.~\ref{sec:bsmeftint}).

\begin{figure*}[!t]
	\subfigure[\label{fig:mu_vs_mhp}]{\includegraphics[width=0.49\textwidth]{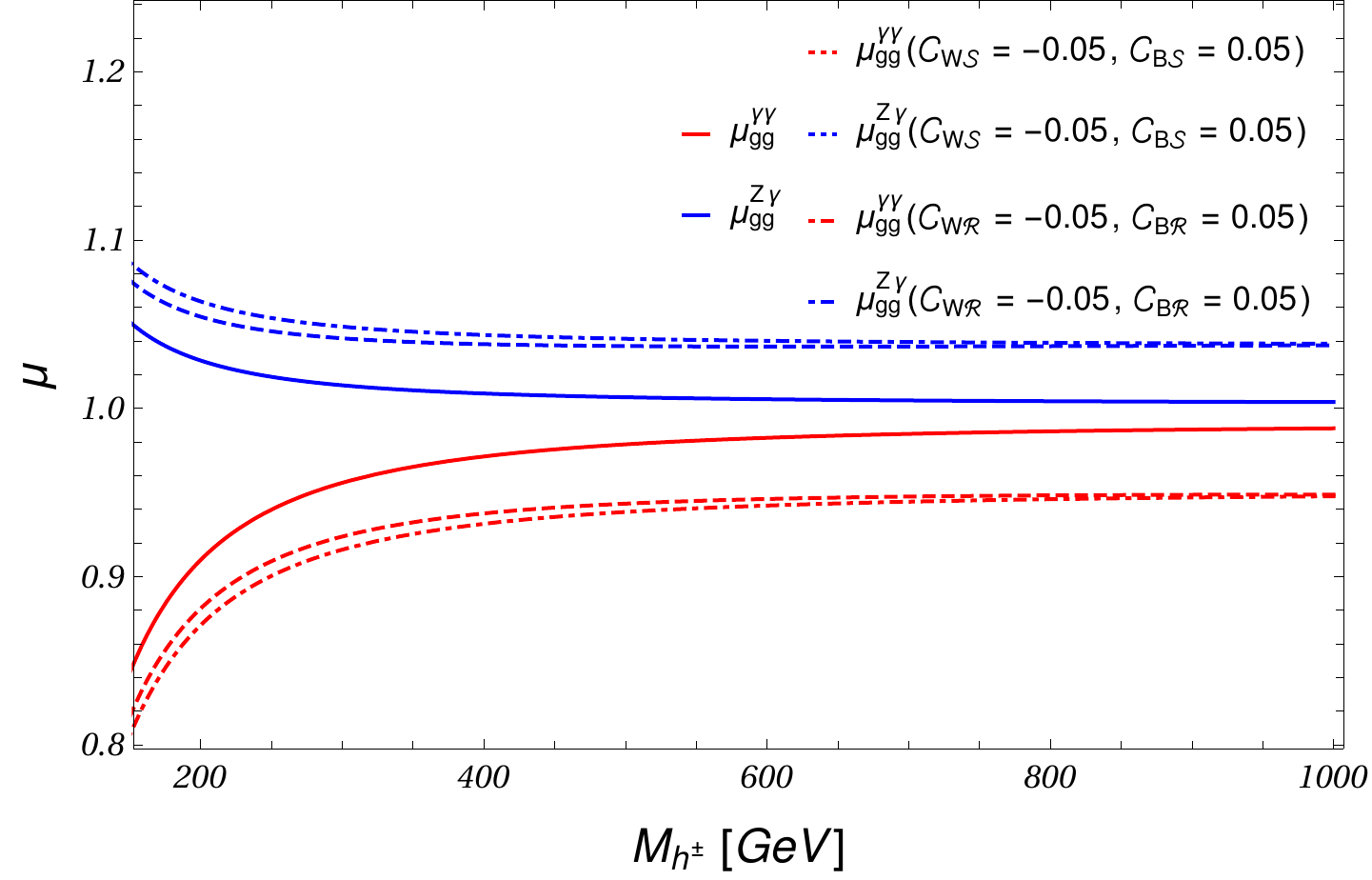}}\hfill
	\subfigure[\label{fig:mu_vs_mhp_5TeV}]{\includegraphics[width=0.49\textwidth]{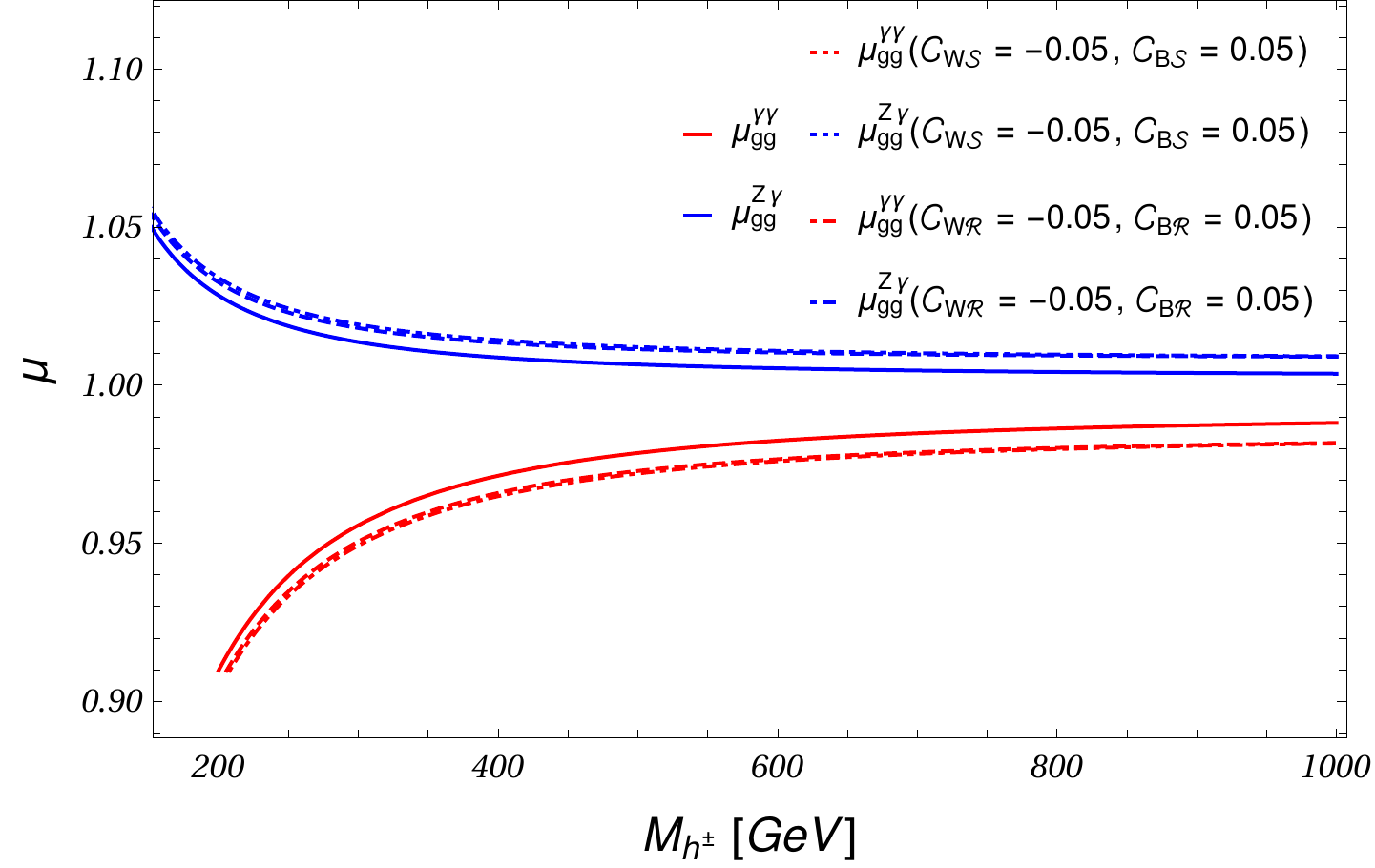}}
	\subfigure[\label{fig:mu_vs_mhpp}]{\includegraphics[width=0.49\textwidth]{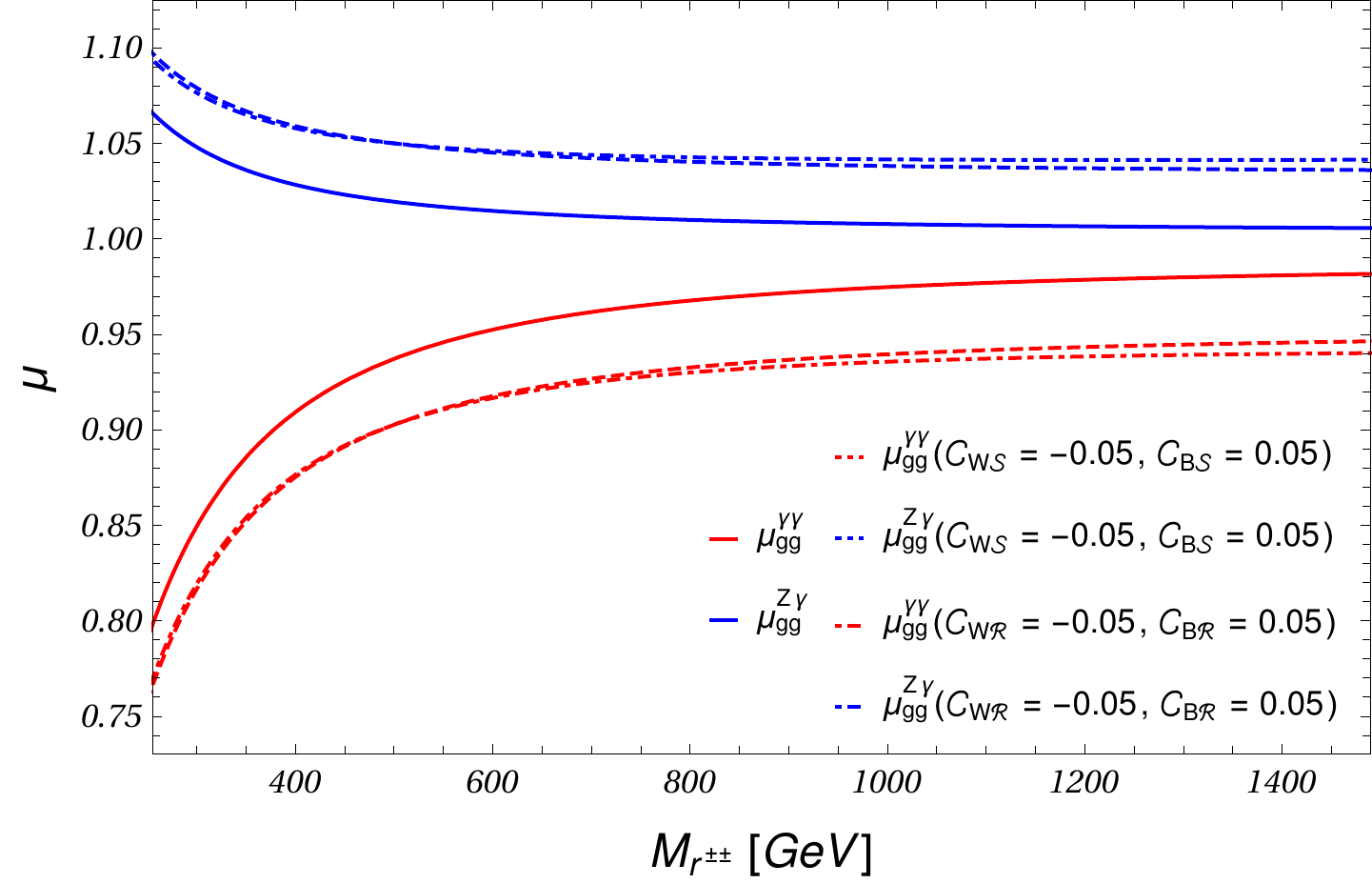}}\hfill
	\subfigure[\label{fig:mu_vs_mhpp_5TeV}]{\includegraphics[width=0.49\textwidth]{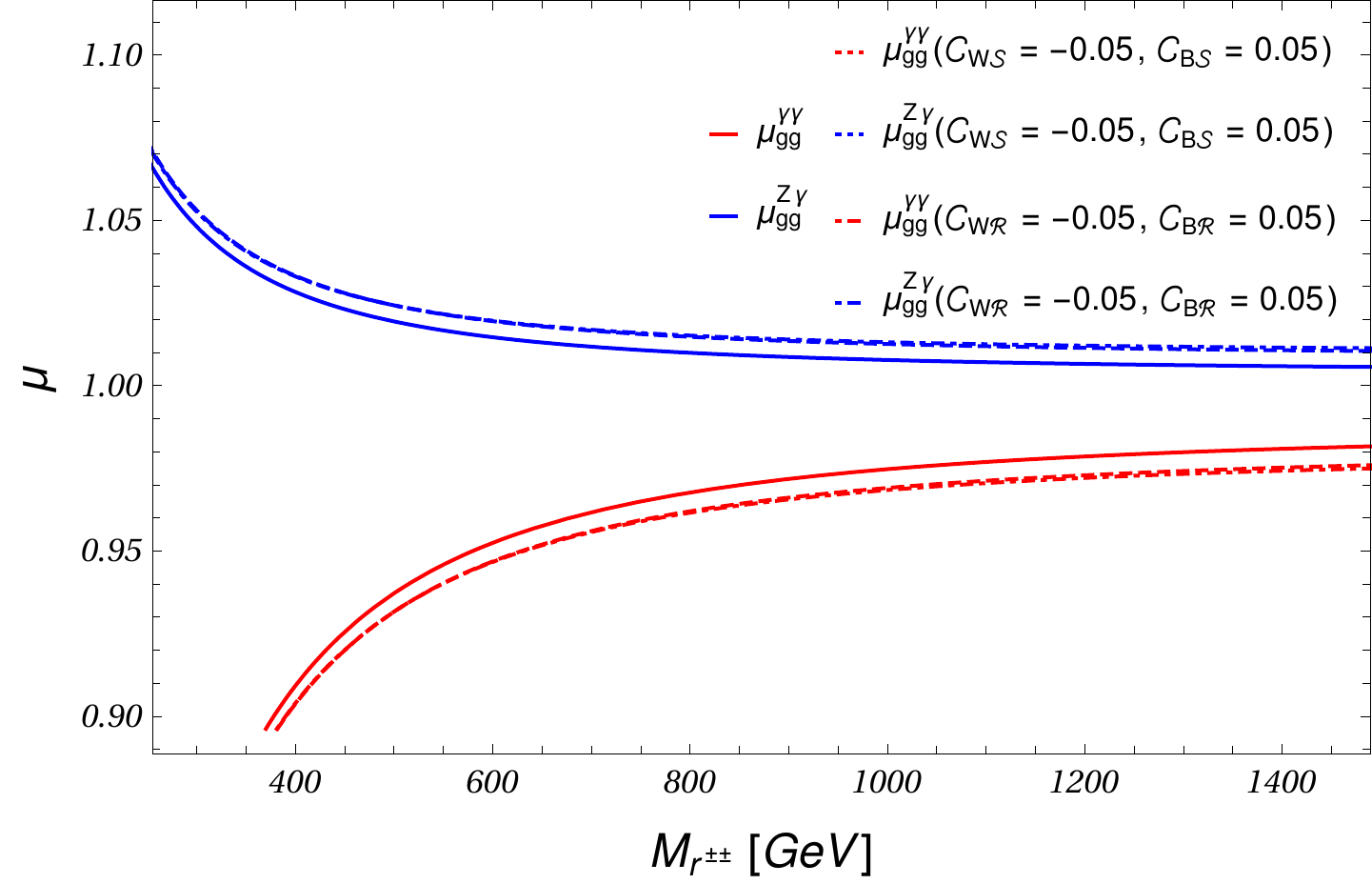}}
	\caption{Impact of representative effective operators on signal strength as a function of singly charged scalar mass $M_{h^{\pm}}$ and doubly charged scalar mass $M_{r^{\pm\pm}}$. For (a) and (b) we consider $M_{r^{\pm\pm}} =1.2$ TeV and for (c) and (d) $M_{h^{\pm}} =0.5$ TeV. For (a) and (c) $\Lambda$ is taken to be 2 TeV, for (b) and (d) $\Lambda=5$ TeV respectively, $\lambda_{4}=\lambda_{5}=1$. \label{fig:Hdata}}
\end{figure*}
	
\subsection{Loop-induced $M_H=125~\text{GeV}$ Higgs physics}
\label{sec:hdec}
We now turn to the discussion of the impact of the model discussed in Sec.~\ref{sec:model} on the loop-induced phenomenology of the 125 GeV Higgs boson. Assuming the narrow width approximation (NWA), we consider the signal strengths from dominant gluon fusion production~\cite{LHCHiggsCrossSectionWorkingGroup:2011wcg} (see also \cite{Georgi:1977gs,Dawson:1990zj,Djouadi:1991tka})
\begin{equation}
	\mu^X_{gg} = {[{\sigma_{\text{GF}} \times \text{BR}(H\to X) ]^\text{BSM}} \over
		[{\sigma_{\text{GF}} \times \text{BR}(H\to X) ]^\text{SM}} }\,.
	\end{equation}
	The CMS experiment predict~\cite{CMS:2017cwx} a sensitivity in the experimentally clean $H\to \gamma\gamma$ channel of
	\begin{equation}
	\label{eq:haa}
	{\Delta\mu^{\gamma\gamma}_{gg} \over \mu^{\gamma\gamma}_{gg}} = 3.3\%
	\end{equation}
	at a (HL-)LHC luminosity of 3/ab. Sensitivity in the $Z\gamma$ channel has been considered in~\cite{deBlas:2019rxi} (for a recent analysis see \cite{ATLAS:2020qcv}) providing a HL-LHC estimate of 
	\begin{equation}
	{\Delta\mu^{Z\gamma}_{gg}\over\mu^{Z\gamma}_{gg}} = 18\%\,.
\end{equation} 
Mapping these sensitivity intervals onto BSM-modified SM predictions, we include the effective interactions of Sec.~\ref{sec:model} to $H\to gg$ (which relates to Higgs production via unitarity~\cite{Djouadi:2005gi}), and $H\to Z\gamma$, as well as $H\to \gamma \gamma$. This leads to one-loop sensitivity to the operators listed in Tab.~\ref{table:Higgs-decay}. Similar to our discussion in Sec.~\ref{sec:gm2}, the inclusion of BSMEFT interactions leads to a renormalisation of the SMEFT counterparts as outlined in Ref.~\cite{Anisha:2021fzf}.
	
	\begin{table}[!b]
		\centering
		\renewcommand{\arraystretch}{1.9}
		{\scriptsize\begin{tabular}{||c|c|c||}
				\hline
				\hline
				\multirow{2}{*}{Decay mode} & Renormalisable & Contributing \\
				& couplings & operators\\
				\hline
				\hline
				\multirow{4}{*}{$H \to \gamma \gamma$} & \multirow{4}{*}{$\lambda_{4},\lambda_{5}$} & $\mathcal{O}_{\phi \mathcal{R}\mathcal{D}}$,\hspace{2mm}$\mathcal{O}_{\phi \mathcal{R}}$,\hspace{2mm}$\mathcal{O}_{\phi \mathcal{S}}$,\\
				& & $\mathcal{O}_{\phi \mathcal{S}\mathcal{D}}$,\hspace{2mm}$\mathcal{O}_{ B \mathcal{R}}$,\hspace{2mm}$\mathcal{O}_{\widetilde{B} \mathcal{R}}$,\\
				& & $\mathcal{O}_{ W \mathcal{R}}$,\hspace{2mm}$\mathcal{O}_{ \widetilde{W} \mathcal{R} }$,\hspace{2mm}$\mathcal{O}_{ B \mathcal{S}}$,\\
				& &$\mathcal{O}_{\widetilde{B} \mathcal{S}}$,\hspace{2mm}$\mathcal{O}_{ W \mathcal{S}}$,\hspace{2mm}$\mathcal{O}_{ \widetilde{W} \mathcal{S} }$.\\
				\hline
				\multirow{5}{*}{$H \to Z \gamma$} & \multirow{5}{*}{$\lambda_{4},\lambda_{5}$} & $\mathcal{O}_{\phi \mathcal{R}\mathcal{D}}$,\hspace{2mm}$\mathcal{O}_{\mathcal{R} \phi \mathcal{D}}$,\hspace{2mm}$\mathcal{O}_{\phi \mathcal{R}}$,\\
				& &$\mathcal{O}_{\phi \mathcal{S}}$,\hspace{2mm}$\mathcal{O}_{\phi \mathcal{S}\mathcal{D}}$,\hspace{2mm}$\mathcal{O}_{\mathcal{S} \phi \mathcal{D}}$,\\
				& &$\mathcal{O}_{ B \mathcal{R}} $,\hspace{2mm}$\mathcal{O}_{\widetilde{B} \mathcal{R}}$,\hspace{2mm}$\mathcal{O}_{ W \mathcal{R}}$,\\
				& &$\mathcal{O}_{ \widetilde{W} \mathcal{R} }$,\hspace{2mm}$\mathcal{O}_{ B \mathcal{S}}$,\hspace{2mm}$\mathcal{O}_{\widetilde{B} \mathcal{S}}$,\\
				& &$\mathcal{O}_{ W \mathcal{S}}$,\hspace{2mm}$\mathcal{O}_{ \widetilde{W} \mathcal{S} }$.\\	
				\hline
				\multirow{2}{*}{$H \to g g$} & \multirow{2}{*}{$\lambda_{4},\lambda_{5}$} & $\mathcal{O}_{G\mathcal{R}}$,\hspace{2mm}$\mathcal{O}_{\widetilde{G}\mathcal{R}}$,\\
				& &$\mathcal{O}_{G\mathcal{S}}$,\hspace{2mm}$\mathcal{O}_{\widetilde{G}\mathcal{S}}$.\\
				\hline
		\end{tabular}}
		\caption{The parameters and the singly- and doubly-charged scalar related operators which contribute to the corrections in prominent loop-induced $H$-decay modes.}
		\label{table:Higgs-decay}
	\end{table}
	
In Fig.~\ref{fig:Hdata}, we demonstrate the impact of the scenario of this paper on the considered Higgs signal strength measurements. While the charged scalars modify the $H\to Z\gamma,\gamma \gamma$ branchings via their hypercharge quantum numbers, their effective operator interactions can lead to significant modifications of the branching, in particular when these relate to the electroweak gauge interactions. We find that $H\to \gamma \gamma$ decays typically provide more stringent limits than $H\to Z\gamma$.

\begin{figure}[!t]
	\includegraphics[width=0.48\textwidth]{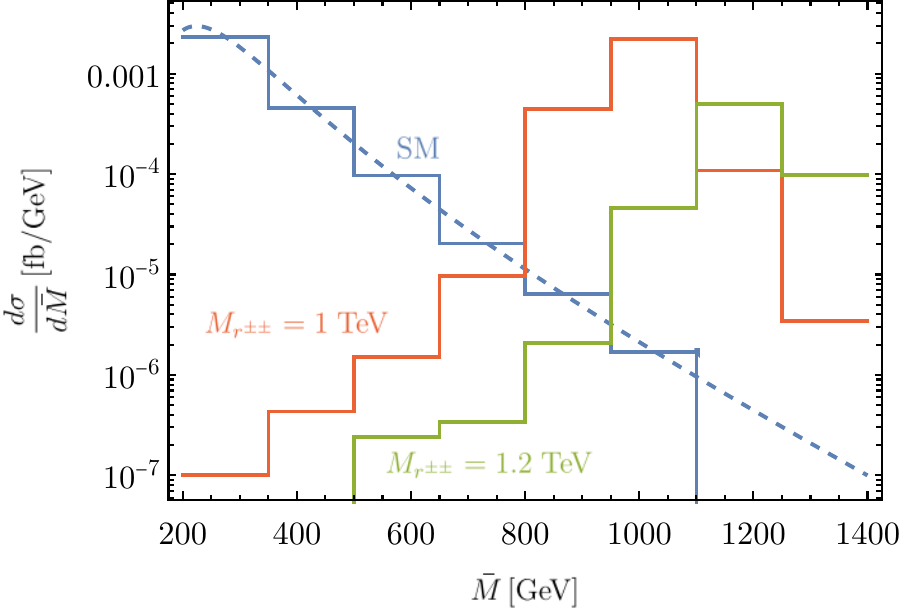}
	\caption{$\Mbar$ distribution for the SM and new physics contributions. The dashed line shows the fit of the background and for the signal $f_{\mathcal{R}} = 0.1$ and a representative $\bar{C}_{\phi \mathcal{R} \mathcal{D}} = 0.3$ was used with the remaining BSMEFT WCs set to zero. The mass of the singly charged scalar is set to $480$ ($500$) GeV for the scenario with $M_{r^{\pm\pm}} = 1$~TeV ($M_{r^{\pm\pm}} = 1.2$~TeV). \label{fig:example_histo}}
\end{figure}

\subsection{Direct LHC sensitivity to doubly charged scalars}
\label{sec:hppdec}
The scalars of Sec.~\ref{sec:model} can be produced at colliders via their hypercharge quantum numbers, implying a predominant production in pairs via Drell-Yan like processes, which is common to many charged scalar extensions of the SM (see e.g.~\cite{Boos:2018fnt,BergeaasKuutmann:2017yud,Han:2007bk}). The production of two $r^{\pm\pm}$ is more efficient than pair production of $h^\pm$ due to its larger charge when assuming similar masses. It will also dominate over $r^{\pm \pm}r^\ast$ along with $r^\ast \to h^\mp h^\mp$ though a virtual $r^*$, see Ref.~\cite{Nebot:2007bc}. The $r^{\+\+}$ decay phenomenology that we will consider in more detail in this section is characterised by decays $r^{\+\+} \to h^\+ h^\+$ (when kinematically accessible)
\begin{multline}
	\Gamma(r^{\pm\pm}\to h^\pm h^\pm) = {\beta \over 128 \pi M_{r^{\pm\pm}}}\\ \times\left\{ 2 {{\cal{C}}_r\over \Lambda} \,v^2 + m \left( {{\cal{C}}_{\phi \mathcal{R}\mathcal{D}} \over \Lambda^2}\,v^2 + 2 \,{ {\cal{C}}_{\phi \mathcal{S}\mathcal{D}}\over \Lambda^2}\,v^2 - 4 \right)\right\}^2 \,,
\end{multline}
where $\beta$ is the $h^\pm$ velocity in the $r^{\pm\pm}$ rest frame, as well as same sign lepton decays, e.g. 
\begin{multline}
	\Gamma(r^{\pm\pm}\to \mu^\pm \mu^\pm) = {M_{r^{\pm\pm}} \over 128 \pi} \\
	\times\left\{
	\left(4 f_{\mathcal{R},\mu\mu}- \left[2\, C_{e \mathcal{R} \phi} + f_{\mathcal{R},\mu\mu} {\cal{C}}_{\phi \mathcal{R}\mathcal{D}} \right]{v^2\over \Lambda^2} \right)^2 + 4\, {C_{ l\phi \mathcal{R}}^2\over \Lambda^4}\,v^2
	\right\}
\end{multline}
in the limit $m_{e^+}\ll M_{r^{\+\+}}$.\footnote{This extends the results of, e.g. Ref.~\cite{Herrero-Garcia:2014hfa} to EFT interactions. These results can be straightforwardly linearised in $\sim 1/\Lambda^2$.}

In the Zee-Babu model, pair production of the doubly-charged scalar through Drell-Yan $p p \to Z/A \to r^{++} r^{--}$ is only affected by the values of SM couplings and any change in production rate arises through BSMEFT operators. Focusing on the overlap of contributing operators between Drell-Yan, Higgs decays and anomalous muon magnetic moment, we note that only $\OphiRD$ contributes in the $r^{\pm \pm}$ pair production through a rescaling of the $r$ field. Considering the possible subsequent decays with leptonic final states we anticipate that the experimental sensitivity of channels with decays to $h^\pm$ will be significantly impacted by the presence of neutrinos that appear as missing energy. Additionally, any final state involving tau leptons will yield a decreased sensitivity due to the difficulty in tagging them in detectors compared to muons and electrons. In contrast, the four lepton channel $r^{++} r^{--} \to \ell^+ \ell^+ \ell^- \ell^-$ will provide a clear signature with a suppressed SM background when the fact that $r^{\pm \pm}$ is the only particle in the model decaying to same-charge leptons is exploited in the analysis. 

We model the new physics interactions using {\sc{FeynRules}}~\cite{Christensen:2008py,Alloul:2013bka} and exporting them in the {\sc{Ufo}}~\cite{Degrande:2011ua} format that can be imported in {\sc{MadGraph}}~\cite{Alwall:2014hca}. Events for the Zee-Babu and BSMEFT are generated with {\sc{MadEvent}}~\cite{Alwall:2011uj,deAquino:2011ub,Alwall:2014hca} including only the interference effects of $\OphiRD$. We include all SM processes contributing to $p p \to \ell^+ \ell^- \ell^+ \ell^-$ as background with a generation-level cut vetoing events in the $M_Z \pm 3.5\, \Gamma_Z$ interval, where $M_Z$ and $\Gamma_Z$ are the invariant mass and decay width of the (virtual) $Z$ boson, respectively. Total decay widths for the charged scalars are calculated with {\sc{MadWidth}}~\cite{Alwall:2014bza}, and cross checked against our analytical results. The events are generated with a fixed branching ratio $\text{BR}(r^{\pm\pm} \to \ell^\pm \ell^\pm)$ and we subsequently rescale the rates under the assumption of the NWA. 

\begin{figure*}[!t]
		\includegraphics[width=0.7\textwidth]{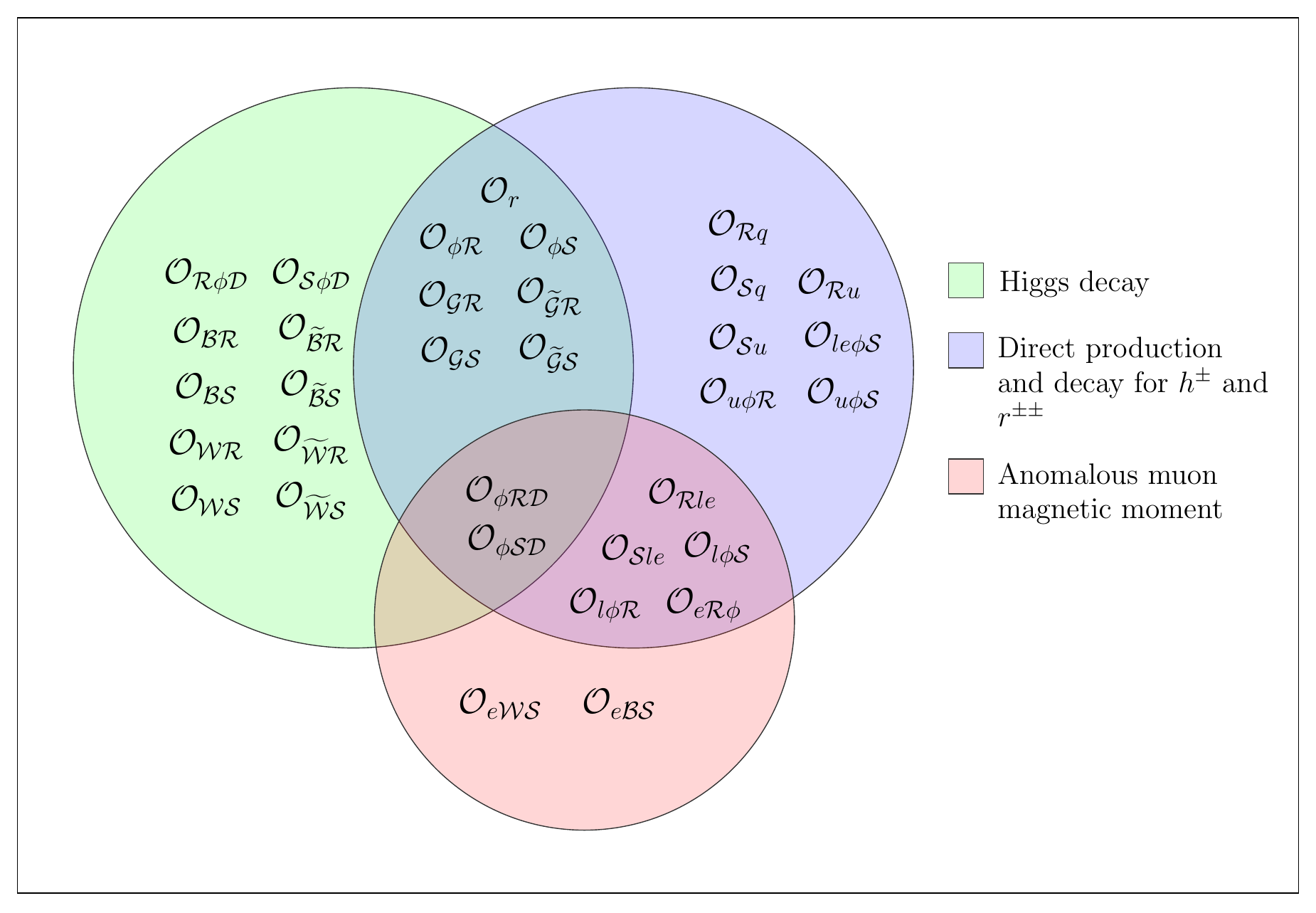}
		\caption{Diagram depicting BSMEFT operators that contribute to three measurements considered in the calculation. $\mathcal{O}_{r},\mathcal{O}_{\phi\mathcal{R}}, \mathcal{O}_{\phi\mathcal{S}}, \mathcal{O}_{\mathcal{G}\mathcal{R}}, \mathcal{O}_{\widetilde{\mathcal{G}}\mathcal{R}}, \mathcal{O}_{\mathcal{G}\mathcal{S}}, \mathcal{O}_{\widetilde{\mathcal{G}}\mathcal{S}}$ are the common operators contributing to both Higgs-decay and processes relevant for direct detection for production and decay for charged scalars. $\mathcal{O}_{\mathcal{R}le},\mathcal{O}_{\mathcal{S}le},\mathcal{O}_{l\phi\mathcal{S}},\mathcal{O}_{l\phi\mathcal{R}},\mathcal{O}_{e\mathcal{R}\phi}$ contribute to anomalous muon magnetic moment as well as charged scalar production and decay processes. $\mathcal{O}_{\phi\mathcal{R}\mathcal{D}},\mathcal{O}_{\phi\mathcal{S}\mathcal{D}}$ contribute to all three processes. A range of the operators are mass-suppressed thus leading to a small overlap in the limit of vanishing quark/lepton masses (e.g. when considering the parton model of LHC collisions).}\label{fig:common_ops}
\end{figure*}

Our analysis is based on the ATLAS search for doubly charged scalars in Ref.~\cite{ATLAS:2017xqs} with relaxed cuts and is performed at parton-level to obtain a qualitative, proof-of-principle comparison. Selection of our analysis requires that all light leptons are in the central part of the detector ($\abs{\eta(\ell)} < 2.5$) with a transverse momentum of $p_T(\ell) > 30$~GeV. Only leptons with no jet activity within the cone radius $\Delta R (j, \ell) = \sqrt{\Delta \eta^2 + \Delta \phi} < 0.4$ are considered and we require exactly four leptons with one positively-charged pair and one negatively-charged, otherwise the event is vetoed (we do not include charge mis-tagging or other experimental systematic uncertainties). A cut is imposed on the invariant mass of each pair such that $m_{\ell^\pm \ell^\pm} > 200$~GeV always. Since the same-charged leptons must be a result of $r^{\pm\pm}$ decays we check the consistency of the two masses by calculating 
\begin{equation}
\Mbar = {m_{\ell^+ \ell^+} + m_{\ell^- \ell^-}\over 2} \,,
\end{equation}
and 
\begin{equation}
\Delta M = \abs{m_{\ell^+ \ell^+} - m_{\ell^- \ell^-}}\,.
\end{equation}
 The two invariant masses are considered consistent if $\Delta M / \Mbar < 0.25$ is satisfied, thus imposing the resonant signal character. Finally, the event is vetoed if a same-flavour, oppositely-charged pair exists with invariant mass in the interval $m_{\ell^+ \ell^-} \in \left[81.2, 101.2\right]$ GeV in order to suppress any background resulting from decays of $Z$ bosons.

We evaluate the sensitivity of LHC using events measure from the $\overline{M}$ differential distribution. Including the new physics contributions, the distribution is given by 
\begin{equation}
	\label{eq:diffdist}
	\frac{\text{d}\sigma}{\text{d}\Mbar} = \frac{\text{d}\sigma_{\text{SM}}}{\text{d}\Mbar} + \frac{\text{d}\sigma_{\text{BZ}}}{\text{d}\Mbar} + \frac{\CphiRD}{\Lambda^2} \frac{\text{d}\sigma_{\phi \mathcal{R} \mathcal{D}}}{\text{d}\Mbar}\;,
\end{equation}
where $\sigma_{\text{SM}}$ denotes the Standard Model contribution and $\sigma_{\text{BZ}} = \sigma_{\text{BZ}}(f_{\mathcal{R}}, M_{r^{\pm\pm}}, M_{h^{\pm}})$ the pure Zee-Babu, which depends on the $f_\mathcal{R}$ coupling and the masses of $r^{\pm\pm}$ and $h^\pm$. The dimension-6 interference contribution from $\OphiRD$ is denoted as $\sigma_{\phi \mathcal{R} \mathcal{D}}$ and also depends on the same parameters as $\sigma_{\text{BZ}}$. The new physics contributions are rescaled with a K-factor value of $1.3$ (see e.g.~\cite{Altarelli:1979ub}) to include higher order corrections. We note that the dependence on $f_\mathcal{R}$ and $M_h^{\pm}$ enters through the branching ratio $\text{BR}(r^{\pm\pm} \to \ell^\pm \ell^\pm)$ and the ratio's dependence on $\CphiRD$ cancels out when no other BSMEFT operator is included. This allows us to obtain contributions for different values of $f_\mathcal{R}$ by rescaling assuming the NWA and to generate events for interference effects caused by $\OphiRD$ independent of $\CphiRD$.

The $\Mbar$ distribution obtained from SM processes is fitted away from the signal region 
to obtain an experimentally-driven estimate for the background for large values of $\Mbar$. The $\Mbar$ distribution for particular values of new physics parameters is shown in Fig.~\ref{fig:example_histo}. We evaluate the signal and background number of events in the region $\Mbar> 200$~GeV at an integrated luminosity of $3$/ab as $S$ and $B$, respectively and calculate, the significance $S/\sqrt{B}$ ($S/\sqrt{S+B}$) under the SM (new physics) hypothesis. We will comment on the search's sensitivity in the next section.

\section{BSMEFT Interplay}
\label{sec:bsmeftint}
We are now ready to consider the phenomenological interplay of the observables discussed in the previous Sec.~\ref{sec:pheno}.\footnote{As indicated by the renormalisation procedure, the measurements considered in this work would be part of the input data in comprehensive global fit. In this work we limit ourselves to the phenomenological interplay of the three measurement methodologies assuming vanishing SMEFT contributions.} In Fig.~\ref{fig:common_ops}, the Venn diagram shows the common operators contributing to all three processes discussed in Sec.~\ref{sec:pheno}. A number of these operators contribute in a fermion mass-suppressed way. The dominant overlap of Higgs data, $a_\mu$ and Drell-Yan production is therefore a single operator $\sim \CphiRD$, which only affects the total with of the exotic scalar search. It is worthwhile to stress that when we do not consider effective interactions related to $h^\pm$, the anomalous magnetic moment is predictive at ${\cal{O}}(\Lambda^{-2})$, i.e. the $r^{\pm\pm}$ contribution to $\Delta a$ is finite even when EFT insertions are considered (see Eq.~\eqref{eq:d6amu}). The interplay of Higgs data, direct sensitivity in LHC searches and anomalous magnetic moment is therefore relatively transparent in the scenario of Sec.~\ref{sec:model}.
\begin{figure*}[!t]
	\subfigure[\label{fig:amufr2}]{\includegraphics[width=0.48\textwidth]{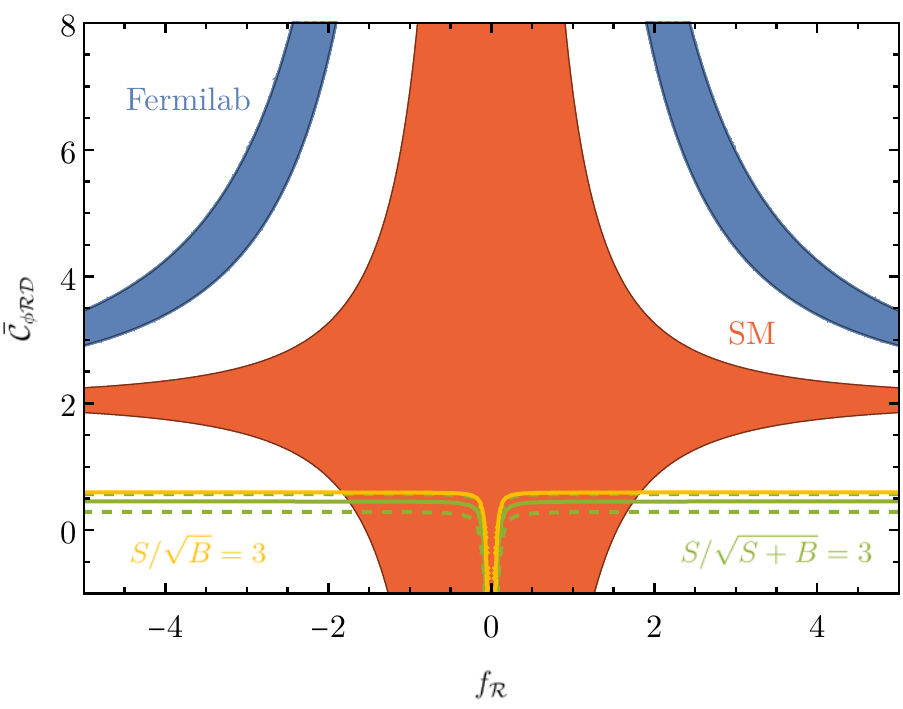}}\hfill
	\subfigure[\label{fig:amufs2}]{\includegraphics[width=0.48\textwidth]{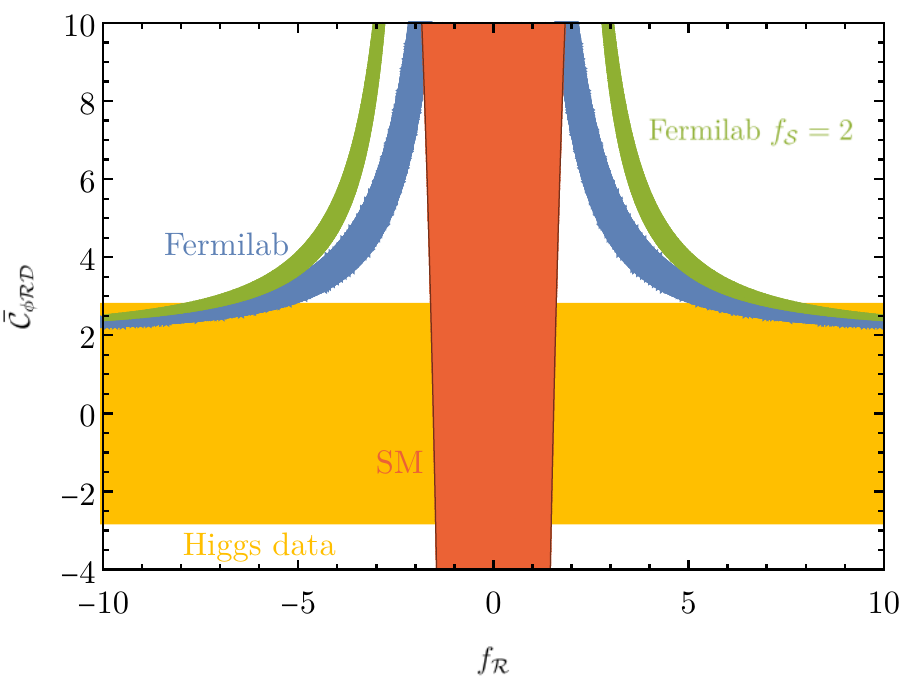}}
	\caption{Regions in the $\fR$-$\CphiRDbar$ plane, where $\bar{\mathcal{C}}_i = \mathcal{C}_i v^2 / \Lambda^2$. Blue and red show the parameter regions where $a_\mu$ is in agreement with the Fermilab experimental measurement and the SM expectation, respectively. On the left figure we set $M_{r^{\pm\pm}} = 1$~TeV and $M_{h^{\pm}} = 0.48$~TeV and also show the $S/\sqrt{B} = 3$ ($S/\sqrt{S+B} = 3$) contour for the direct detection analysis with yellow (green) using a value of $m = 246$~GeV. Note that strong EFT coupling (BSMEFT $>$ BSM) is $|{\bar{C}}_{\phi \mathcal{R} \mathcal{D}}|\gtrsim 0.8$. On the right, we restrict $\CphiSDbar \approx - 4 \CphiRDbar$ and set $M_{r^{\pm\pm}} = M_{h^{\pm}} = 1$~TeV. The contours from the anomalous magnetic moment depend on $\fS$ with blue and red contours shown with a value of zero. We note that any value of $\fS$ smaller than unity does not significantly affect the results. Additionally, we show the contours that yield agreement with the Fermilab measurement when $\fS = 2$ with green and overlay the results for the Higgs decay sensitivity from Eq.~\eqref{eq:haa} for the choice of Eq.~\eqref{eq:cancel}. 
   \label{fig:amu_direct}}
\end{figure*}

In Fig.~\ref{fig:amu_direct}, we show the interplay of the direct search outlined in Sec.~\ref{sec:hppdec} with the anomalous magnetic moment for a particular mass choice of the exotic charged scalars (including open decays $r^{\pm\pm}\to h^\pm h^\pm$). The blue contour refers to the Fermilab $a_\mu$ measurement while the red contour shows the SM expectation as provided in~Ref.~\cite{Aoyama:2020ynm}
\begin{equation}
a_\mu(\text{SM}) = (116 591 810 \pm 43) \times 10^{-11}\;,
\end{equation}
when the uncertainty is used as a limit for new physics. The size of the Fermilab/BNL excess can be compensated by contributions that can be attributed to new BSM physics, overcoming the limitations of the renormalisable Zee-Babu model, however, at strong coupling $\CphiRD{ \text{TeV}^2 /\Lambda^2} \simeq 66$. This is due to the fact that the EFT contribution, whilst not being logarithmically enhanced has to overcome the renormalisable contribution of the charged scalars. As already alluded to in Sec.~\ref{sec:gm2}, this can be mitigated by considering charged scalar contributions. Our $r^{\+\+}$-related findings are qualitatively similar to results reported in other model-specific $a_\mu$ analyses~\cite{Baer:2021aax,Athron:2021iuf,Frank:2021nkq,Ellis:2021zmg,Zhang:2021dgl,Jueid:2021avn,Altmannshofer:2021hfu,Chakraborti:2021squ,Chakraborti:2021dli}: BSM states are forced to be light and/or strongly coupled to address the $a_\mu$ anomaly. Including signal extrapolations at the LHC as shown in Fig.~\ref{fig:amu_direct}(a) shows that any evidence for new doubly charged states at the LHC would stand in stark contrast with the $a_\mu$ measurement when interpreted from an extended Zee-Babu perspective.

Including Higgs physics (which is dominated by $\mu^{\gamma\gamma}_{gg}$) leads to further tension. Even when direct renormalisable trilinear $H-r^{++}-r^{--}$ couplings are dialled small $\lambda_{5}\simeq 0$ (note that Eq.~\eqref{eq:7} includes this limit), $\mathcal{O}_{\phi\mathcal{R}\mathcal{D}}$ (see Tab.~\ref{table:ops_structures}) introduces the $r^{\+\+}$ loop contributions to the Higgs signal strength $\mu^{\gamma\gamma}_{gg}$, which at this point in the LHC programme is already constrained at the 10\% level. Scanning the Higgs signal strength modifications, including the $h^\pm$ interactions and their dimension-6 EFT modifications, we are not able to reconcile SM consistency of the $H\to \gamma\gamma$ branching with the $a_\mu$ anomaly when the latter is attributed to choices in the 
$f_{\mathcal{R}}-{\mathcal{C}}_{\phi\mathcal{R}\mathcal{D}}$ plane. 

\medskip
Opening up the EFT and renormalisable coupling space, cancellations between the charged states and their EFT interactions can appear. This typically requires the full renormalisation of $a_\mu$ as described above. For 
\begin{equation}
\label{eq:cancel}
\mathcal{C}_{\phi\mathcal{S}\mathcal{D}}= - 4\, \mathcal{C}_{\phi\mathcal{R}\mathcal{D}},~ M_{r^{\pm\pm}}\simeq M_{s^{\pm}}\;,
\end{equation} the charged Higgs contributions cancel. The $a_\mu$ excess could then be capture in a mismatch of the Yukawa couplings, see Fig.~\ref{fig:amu_direct}(b) We find that $f_{\mathcal{R}}\sim 5$ and $f_{\mathcal{S}}\sim 1$ can accommodate the Fermilab excess for strong coupling $\bar{\mathcal{C}}_{\phi\mathcal{R}\mathcal{D}} \sim 3$, which implies a $r^{\+\+}$ partial width into a single lepton combination of around $60~\text{GeV}$. Such a state can fall into the LHC kinematic coverage, see Fig.~\ref{fig:amu_direct}(a) and Ref.~\cite{ATLAS:2017xqs}. The further exploration of the high mass doubly charged scalar production is therefore highly motivated in the light of a SM-like Higgs and the consolidated $a_\mu$ anomaly.

\section{Conclusions}
\label{sec:conc}
The recent Fermilab consolidation of $a_\mu$ raises the question of how new physics can be accommodated as the exotics and Higgs precision programme 
is evolving at the LHC. The direct sensitivity at the LHC with its so far null results in BSM searches moves new physics scales into regions where it becomes challenging to accommodate a significant anomalous magnetic moment of the muon when we take the BNL/Fermilab results as indication for BSM interactions. In this work we have approached the interplay of these experimental arenas by means of effective field theory. A significant muon magnetic moment requires the presence of relatively light charged degrees of freedom which we supplement with a complete dimension-5 and -6 effective field theory analysis. The Zee-Babu scenario as a particularly motivated BSM candidate theory gives then rise to a range of BSMEFT interactions that enable us to discuss $a_\mu$ precision results in tension with expected developments at the LHC.

Obviously, the rather large number of relevant Wilson coefficients exceed the number of measurements that result from Higgs physics, $a_\mu$ and direct sensitivity via $r^{\+\+}$ pair production, yet the overlapping set of operators that simultaneously affects all measurements and searches is relatively small and shows a significant tension when the SM expectation for Higgs physics is considered: Agreement of $a_\mu$ requires a significant deformation of charged scalar interactions, which in turn highly modify Higgs physics beyond experimentally allowed constraints. While this is particularly pronounced when we limit ourselves to the $r^{\+\+}$ state, we find that the additional freedom provided by the EFT extension of the $h^\+$ interactions can be exploited to achieve cancellations that render Higgs data compatible with the SM observation whilst obtaining a sizeable $a_\mu$, again at relatively large couplings. On the one hand, this provides an important constraint for potential UV completions that the considered scenarios seeks to inform. On the other hand, the associated parameter ranges can be explored by future searches for doubly charged scalar states as done in e.g.~\cite{ATLAS:2020qcv}. Additional constraints can in principle be resolved in more challenging $pp\to 2\ell/4\ell +\slashed{E}_T$ searches, the impact of which we leave for future work.

\acknowledgements	
The work of A, U.B., and J.C. is supported by the Science and Engineering Research Board, Government
of India, under the agreement SERB/PHY/2019501 (MATRICS). 
C.E. is supported by the UK Science and Technology Facilities Council (STFC) under grant ST/T000945/1 and by the IPPP Associateship Scheme. M.S. is supported by the STFC under grant ST/P001246/1. P.S. is funded by an STFC studentship under grant ST/T506102/1.

\allowdisplaybreaks

\appendix
\section{BSM effective operators}\label{sec:App_A}	
		
Throughout the paper we assume that the charged scalars $\mathcal{S}$ and $\mathcal{R}$ are light enough to be considered as infrared degrees of freedom. Thus after integrating out the new physics at $\Lambda$, we are left with effective operators that lead to the modifications of SM interactions (i.e. the SMEFT operators), or can alter the existent BSM interactions at the renormalisable level, and are hereby identified as BSMEFT operators.
	
The general way to capture the effect of all possible such modifications is to construct a complete and exhaustive set of BSMEFT operators at each mass dimension following Bottom-up approach. A number of models where SM is extended by new degrees of freedom have been discussed in Ref.~\cite{Banerjee:2020jun}.
	
We generated Warsaw-like operator bases of dimensions-5 and -6 operators for the case of Zee-Babu model with \textbf{GrIP} \cite{Banerjee:2020bym}, the explicit structures of these operators have been tabulated below in Tab.~\ref{table:SM+SinglyChargedScalar-dim5-ops-1}, \ref{table:SM+SinglyChargedScalar-dim6-ops-1} and \ref{table:SM+SinglyChargedScalar-dim6-ops-2}. Notably, we obtain a new class of dimension-5 operators $\Phi^5$, which unlike for the SMEFT case, arise due to possible gauge invariant structures allowed by hypercharge quantum numbers of the charged scalars.
\begin{table}[!t]
	\centering
	\renewcommand{\arraystretch}{1.9}
	{\scriptsize\begin{tabular}{||c|c||c|c||}
			\hline
			\hline
			\multicolumn{4}{||c||}{$\Phi^5$}\\
			\hline
			$\mathcal{O}_{r}$&
			$\boldsymbol{(\phi^{\dagger}\,\phi)\,\mathcal{R}^{\dagger}\,\mathcal{S}^2}$&
			$\mathcal{O}_{sr} $&
			$\boldsymbol{(\mathcal{S}^{\dagger}\,\mathcal{S})\,\mathcal{R}^{\dagger}\,\mathcal{S}^2}$\\
			
			$\mathcal{O}_{rs} $&
			$\boldsymbol{(\mathcal{R}^{\dagger}\,\mathcal{R})\,\mathcal{R}^{\dagger}\,\mathcal{S}^2}$&
			&
			\\
			\hline
			\hline
			\multicolumn{4}{||c||}{$\Psi^2\Phi^2$}\\
			\hline
			$\tilde{\mathcal{O}}_{dq\phi\mathcal{S}} $&
			$\boldsymbol{(\overline{Q} \,d) \,(\tilde{\phi} \,\mathcal{S})}$&
			$\tilde{\mathcal{O}}_{uq\phi\mathcal{S}} $&
			$\boldsymbol{(\overline{Q} \,u) \,(\phi \,\mathcal{S})}$\\
			
			$\tilde{\mathcal{O}}_{le\phi\mathcal{S}} $&
			$\boldsymbol{(\overline{L} \,e) \,(\tilde{\phi} \,\mathcal{S})}$&
			$\tilde{\mathcal{O}}_{e\mathcal{S}} $&
			$\boldsymbol{(\overline{e^{c}} \,e) \,\mathcal{S}^2}$\\
			\hline
	\end{tabular}}
	\caption{Explicit structures of the dimension-5 effective operators for the Zee-Babu model. The operators written in bold have distinct hermitian conjugates.}
	\label{table:SM+SinglyChargedScalar-dim5-ops-1}
\end{table}

\begin{table}[!t]
	\centering
	\renewcommand{\arraystretch}{1.9}
	{\scriptsize\begin{tabular}{||c|c||c|c||}
			\hline
			\hline
			\multicolumn{2}{||c||}{$\Phi^6$}&
			\multicolumn{2}{c||}{$\Phi^4\mathcal{D}^2$}
			\\
			\hline
			$\mathcal{O}_{\mathcal{S}} $&
			$(\mathcal{S}^{\dagger} \,\mathcal{S})^3$&
			$\mathcal{O}_{\mathcal{S}\square} $&
			$(\mathcal{S}^{\dagger} \,\mathcal{S}) \,\square \,(\mathcal{S}^{\dagger} \,\mathcal{S})$\\
			
			$\mathcal{O}_{\mathcal{S}\phi} $&
			$(\phi^{\dagger} \,\phi) \,(\mathcal{S}^{\dagger} \,\mathcal{S})^2$&
			$\mathcal{O}_{ \mathcal{S}\phi \mathcal{D}} $&
			$(\mathcal{S}^{\dagger}\,\mathcal{S})\,\left[(\mathcal{D}^{\mu}\,\phi)^{\dagger}(\mathcal{D}_{\mu}\,\phi)\right]$\\
			
			$\mathcal{O}_{\phi\mathcal{S}} $&
			$(\phi^{\dagger} \,\phi)^2 \,(\mathcal{S}^{\dagger} \,\mathcal{S})$&
			$\mathcal{O}_{\phi \mathcal{S} \mathcal{D}}$&
			$(\phi^{\dagger}\,\phi)\,\left[(\mathcal{D}^{\mu}\,\mathcal{S})^{\dagger}(\mathcal{D}_{\mu}\,\mathcal{S})\right]$\\
			
			$\mathcal{O}_{\mathcal{R}} $&
			$(\mathcal{R}^{\dagger} \,\mathcal{R})^3$&
			$\mathcal{O}_{\mathcal{R}\square} $&
			$(\mathcal{R}^{\dagger} \,\mathcal{R}) \,\square \,(\mathcal{R}^{\dagger} \,\mathcal{R})$\\
			
			$\mathcal{O}_{\mathcal{R}\phi} $&
			$(\phi^{\dagger} \,\phi) \,(\mathcal{R}^{\dagger} \,\mathcal{R})^2$&
			$\mathcal{O}_{\mathcal{R}\phi \mathcal{D}} $&
			$(\mathcal{R}^{\dagger}\,\mathcal{R})\,\left[(\mathcal{D}^{\mu}\,\phi)^{\dagger}(\mathcal{D}_{\mu}\,\phi)\right]$
			\\
			
			$\mathcal{O}_{\phi \mathcal{R}} $&
			$(\phi^{\dagger} \,\phi)^2 \,(\mathcal{R}^{\dagger} \,\mathcal{R})$&
			$\mathcal{O}_{\phi \mathcal{R} \mathcal{D}}$&
			$(\phi^{\dagger}\,\phi)\,\left[(\mathcal{D}^{\mu}\,\mathcal{R})^{\dagger}(\mathcal{D}_{\mu}\,\mathcal{R})\right]$
			\\
			
			$\mathcal{O}_{\phi \mathcal{R} \mathcal{S}} $&
			$(\phi^{\dagger} \,\phi) \,(\mathcal{R}^{\dagger} \,\mathcal{R}) \,(\mathcal{S}^{\dagger} \,\mathcal{S})$&
			$\mathcal{O}_{\mathcal{S}\mathcal{R} \mathcal{D}}$&
			$(\mathcal{S}^{\dagger}\,\mathcal{S})\,\left[(\mathcal{D}^{\mu}\,\mathcal{R})^{\dagger}(\mathcal{D}_{\mu}\,\mathcal{R})\right]$
			\\
			
			$\mathcal{O}_{\mathcal{S} \mathcal{R}} $&
			$(\mathcal{S}^{\dagger} \,\mathcal{S})^2 \,(\mathcal{R}^{\dagger} \,\mathcal{R})$&
			$\mathcal{O}_{\mathcal{R}\mathcal{S} \mathcal{D}}$&
			$(\mathcal{R}^{\dagger}\,\mathcal{R})\,\left[(\mathcal{D}^{\mu}\,\mathcal{S})^{\dagger}(\mathcal{D}_{\mu}\,\mathcal{S})\right]$
			\\
			
			$\mathcal{O}_{\mathcal{R} \mathcal{S}}$&
			$(\mathcal{S}^{\dagger} \,\mathcal{S}) \,(\mathcal{R}^{\dagger} \,\mathcal{R})^2$&
			$\mathcal{O}_{\phi\mathcal{S} \mathcal{R} \mathcal{D}}$&
			$\boldsymbol{(\mathcal{S}^{\dagger}\,\mathcal{R}) \,\left[(\mathcal{D}^{\mu}\,\tilde{\phi}^{\dagger})(\mathcal{D}_{\mu}\,\tilde{\phi})\right]}$
			\\
			
			$\mathcal{O}_{\mathcal{S}^2 \mathcal{R}}$&
			$\boldsymbol{(\mathcal{S}\,\mathcal{R}^{\dagger}\,\mathcal{S})^2}$&
			&
			\\
			\hline
			\hline
			\multicolumn{2}{||c||}{$\Phi^2X^2$}&
			\multicolumn{2}{c||}{$\Psi^2\Phi^2\mathcal{D}$}\\
			\hline
			$\mathcal{O}_{ B \mathcal{S}} $&
			$B_{\mu\nu} \,B^{\mu\nu} \,(\mathcal{S}^{\dagger} \,\mathcal{S})$&
			$\mathcal{O}_{ \mathcal{S}q} $&
			$(\,\overline{Q} \,\gamma^{\mu} \,Q) \,(\mathcal{S}^{\dagger} \,i\overleftrightarrow{\mathcal{D}}_{\mu} \,\mathcal{S})$\\
			
			$\mathcal{O}_{B\mathcal{R}} $&
			$B_{\mu\nu} \,B^{\mu\nu} \,(\mathcal{R}^{\dagger} \,\mathcal{R})$&
			$\mathcal{O}_{ \mathcal{R} q} $&
			$(\overline{Q} \,\gamma^{\mu} \,Q) \,(\mathcal{R}^{\dagger} \,i\overleftrightarrow{\mathcal{D}}_{\mu} \,\mathcal{R})$\\
			
			$\mathcal{O}_{\tilde{B} \mathcal{S}} $&
			$\tilde{B}_{\mu\nu} \,B^{\mu\nu} \,(\mathcal{S}^{\dagger} \,\mathcal{S})$&
			$\mathcal{O}_{ \mathcal{S} l} $&
			$(\,\overline{L} \,\gamma^{\mu} \,L) \,(\mathcal{S}^{\dagger} \,i\overleftrightarrow{\mathcal{D}}_{\mu} \,\mathcal{S})$\\
			
			$\mathcal{O}_{\tilde{B} \mathcal{R} } $&
			$\tilde{B}_{\mu\nu} \,B^{\mu\nu} \,(\mathcal{R}^{\dagger} \,\mathcal{R})$&
			$\mathcal{O}_{ \mathcal{R} l} $&
			$(\overline{L} \,\gamma^{\mu} \,L) \,(\mathcal{R}^{\dagger} \,i\overleftrightarrow{\mathcal{D}}_{\mu} \,\mathcal{R})$\\
			
			$\mathcal{O}_{G \mathcal{S}} $&
			$G^{A}_{\mu\nu}\,G^{A\mu\nu}\,(\mathcal{S}^{\dagger} \,\mathcal{S})$&
			$\mathcal{O}_{ \mathcal{S}u } $&
			$(\,\overline{u} \,\gamma^{\mu} \,u) \,(\mathcal{S}^{\dagger} \,i\overleftrightarrow{\mathcal{D}}_{\mu} \,\mathcal{S})$\\
			
			$\mathcal{O}_{G \mathcal{R} } $&
			$G^{A}_{\mu\nu}\,G^{A\mu\nu}\,(\mathcal{R}^{\dagger} \,\mathcal{R})$&
			$\mathcal{O}_{\mathcal{R} u} $&
			$(\overline{u} \,\gamma^{\mu} \,u) \,(\mathcal{R}^{\dagger} \,i\overleftrightarrow{\mathcal{D}}_{\mu} \,\mathcal{R})$\\
			
			$\mathcal{O}_{\tilde{G} \mathcal{S} } $&
			$\tilde{G}^{A}_{\mu\nu}\,G^{A\mu\nu}\,(\mathcal{S}^{\dagger} \,\mathcal{S})$&
			$\mathcal{O}_{ \mathcal{S} d} $&
			$(\,\overline{d} \,\gamma^{\mu} \,d) \,(\mathcal{S}^{\dagger} \,i\overleftrightarrow{\mathcal{D}}_{\mu} \,\mathcal{S})$\\
			
			$\mathcal{O}_{\tilde{G} \mathcal{R} } $&
			$\tilde{G}^{A}_{\mu\nu}\,G^{A\mu\nu}\,(\mathcal{R}^{\dagger} \,\mathcal{R})$&
			$\mathcal{O}_{ \mathcal{R} d} $&
			$(\overline{d} \,\gamma^{\mu} \,d) \,(\mathcal{R}^{\dagger} \,i\overleftrightarrow{\mathcal{D}}_{\mu} \,\mathcal{R})$\\
			
			$\mathcal{O}_{ W \mathcal{S}} $&
			$W^{I}_{\mu\nu} \,W^{I\mu\nu} \,(\mathcal{S}^{\dagger} \,\mathcal{S})$&
			$\mathcal{O}_{ \mathcal{S} e} $&
			$(\,\overline{e} \,\gamma^{\mu} \,e) \,(\mathcal{S}^{\dagger} \,i\overleftrightarrow{\mathcal{D}}_{\mu} \,\mathcal{S})$\\
			
			$\mathcal{O}_{W \mathcal{R} } $&
			$W^{I}_{\mu\nu} \,W^{I\mu\nu} \,(\mathcal{R}^{\dagger} \,\mathcal{R})$&
			$\mathcal{O}_{ \mathcal{R} e} $&
			$(\overline{e} \,\gamma^{\mu} \,e) \,(\mathcal{R}^{\dagger} \,i\overleftrightarrow{\mathcal{D}}_{\mu} \,\mathcal{R})$\\

			$\mathcal{O}_{ \tilde{W} \mathcal{S} } $&
			$\tilde{W}^{I}_{\mu\nu} \,W^{I\mu\nu} \,(\mathcal{S}^{\dagger} \,\mathcal{S})$&
			$\mathcal{O}_{ \mathcal{S} l e} $&
			$\boldsymbol{(\overline{L^c}\,\gamma^{\mu}\,e)(\widetilde{\phi}\,i\mathcal{D}_{\mu}\,\mathcal{S})}$
			\\
			
			$\mathcal{O}_{\tilde{W} \mathcal{R} } $&
			$\tilde{W}^{I}_{\mu\nu} \,W^{I\mu\nu} \,(\mathcal{R}^{\dagger} \,\mathcal{R})$&
			$\mathcal{O}_{\mathcal{R}le}$&
			$\boldsymbol{(\overline{L^c}\,\gamma^{\mu}\,e)(\phi\,i\mathcal{D}_{\mu}\,\mathcal{R})}$
			\\
			\hline
	\end{tabular}}
	\caption{Explicit structures of the dimension-6 effective operators for the Zee-Babu model. The operators written in bold have distinct hermitian conjugates.}\label{table:SM+SinglyChargedScalar-dim6-ops-1}
\end{table}	

\begin{table}[!t]
	\centering
	\renewcommand{\arraystretch}{1.9}
	{\scriptsize\begin{tabular}{||c|c||c|c||}
			\hline
			\hline
			\multicolumn{4}{||c||}{$\Psi^2\Phi^3$}\\
			\hline
			
			$\mathcal{O}_{ e \phi \mathcal{S}} $&
			$\boldsymbol{(\overline{L} \,e) \,\phi \,(\mathcal{S}^{\dagger} \,\mathcal{S})}$&
			$\mathcal{O}_{l \phi \mathcal{S}} $&
			$\boldsymbol{(\overline{L^{c}} \,i\tau_{2}\,L) \,\mathcal{S} \,(\phi^{\dagger} \,\phi)}$\\
			
			$\mathcal{O}_{ u \phi \mathcal{S}} $&
			$\boldsymbol{(\overline{Q} \,u) \,\tilde{\phi} \,(\mathcal{S}^{\dagger} \,\mathcal{S})}$&
			$\mathcal{O}_{l \mathcal{S}} $&
			$\boldsymbol{(\overline{L^{c}} \,i\tau_{2}\,L) \,\mathcal{S} \,(\mathcal{S}^{\dagger} \,\mathcal{S})}$\\
			
			$\mathcal{O}_{d \phi \mathcal{S}} $&
			$\boldsymbol{(\overline{Q} \,d) \,\phi \,(\mathcal{S}^{\dagger} \,\mathcal{S})}$&$\mathcal{O}_{e \mathcal{R}\phi} $&
			$\boldsymbol{(\overline{e^{c}} \,e) \,\mathcal{R} \,(\phi^{\dagger} \,\phi)}$
			\\
			
			$\mathcal{O}_{ e \phi \mathcal{R}} $&
			$\boldsymbol{(\overline{L} \,e) \,\phi \,(\mathcal{R}^{\dagger} \,\mathcal{R})}$&
			$\mathcal{O}_{l \phi \mathcal{R}} $&
			$\boldsymbol{(\overline{L^{c}} \,i\tau_{2}\,L)  \,( \,\phi^{\dagger}\,\mathcal{R}\,\tilde{\phi})}$\\
			
			$\mathcal{O}_{ u \phi \mathcal{R}} $&
			$\boldsymbol{(\overline{Q} \,u) \,\tilde{\phi} \,(\mathcal{R}^{\dagger} \,\mathcal{R})}$&
			$\mathcal{O}_{e \mathcal{R}} $&
			$\boldsymbol{(\overline{e^{c}} \,e) \,\mathcal{R} \,(\mathcal{R}^{\dagger} \,\mathcal{R})}$
			\\
			
			$\mathcal{O}_{ d \phi \mathcal{R}} $&
			$\boldsymbol{(\overline{Q} \,d) \,\phi \,(\mathcal{R}^{\dagger} \,\mathcal{R})}$&
			$\mathcal{O}_{e \mathcal{S} \mathcal{R}} $&
			$\boldsymbol{(\overline{e^{c}} \,e) \,\mathcal{R} \,(\mathcal{S}^{\dagger}\,\mathcal{S})}$
			\\
			
			$\mathcal{O}_{ d \phi \mathcal{R} \mathcal{S}} $&
			$\boldsymbol{(\overline{Q} \,d) \,\tilde{\phi} \,(\mathcal{S}^{\dagger} \,\mathcal{R})}$&
			$\mathcal{O}_{ u \phi \mathcal{R} \mathcal{S}} $&
			$\boldsymbol{(\overline{Q} \,u) \,\phi \,(\mathcal{R}^{\dagger} \,\mathcal{S})}$
			\\
			\hline
			\hline
			\multicolumn{4}{||c||}{$\Psi^2\Phi X$}\\
			\hline
			
			$\mathcal{O}_{e B \mathcal{S}} $&
			$\boldsymbol{B_{\mu\nu} \,(\overline{L^{c}}\,\sigma^{\mu\nu} \,L) \,\mathcal{S}}$&
			$\mathcal{O}_{e W \mathcal{S}} $&
			$\boldsymbol{W^I_{\mu\nu} \,(\overline{L^{c}}\,\tau^I\,\sigma^{\mu\nu} \,L) \,\mathcal{S}}$\\
			
			$\mathcal{O}_{ e B\mathcal{R}} $&
			$\boldsymbol{B_{\mu\nu} \,(\overline{e^{c}}\,\sigma^{\mu\nu} \,e) \,\mathcal{R}}$&
			&\\		
			\hline
	\end{tabular}}
	\caption{Explicit structures of the dimension-6 effective operators for the Zee-Babu model. The operators written in bold have distinct hermitian conjugates.}\label{table:SM+SinglyChargedScalar-dim6-ops-2}
\end{table}

\section{Renormalisation}   
\label{app:renorm}
In the following section, we provide the expressions for the renormalisation constants (RC) which we include in the counter terms considered to remove the divergences for the calculation of muon anomalous magnetic moment and Higgs decay to di-photon, as discussed in Sec.~\ref{sec:gm2} and Sec.~\ref{sec:hdec} respectively. For both scenarios, we choose on-shell renormalisation for propagating fields and parameters and $\overline{\text{MS}}$ renormalisation for the Wilson coefficients. The notations used here to express the one-point and the two-point integrals are given along the lines of Ref.~\cite{Anisha:2021fzf}
   \begin{equation}
   \begin{split}
   A_{0}(m^2)&=m^2 \,\Delta+\mathcal{O}(1),\nonumber\\
   B_{0}&=\Delta+\mathcal{O}(1),\nonumber\\
   B_{1}&=-\frac{\Delta}{2}+\mathcal{O}(1),\nonumber\\
   B_{00}(p^2,m_{1}^2,m_{2}^2)&=\left(\frac{m_1^2+m_2^2}{4}-\frac{p^2}{12}\right)\Delta+\mathcal{O}(1), 
   \end{split}
   \end{equation} 
where $\Delta$ denotes the UV-divergent parts associated with the loop integrals which we remove via the renormalisation of the Wilson coefficients (see e.g.~\cite{Denner:1991kt,Denner:2019vbn}. The functions $dB_i$ represent the derivative of the scalar functions $B_i$ with respect to the external momentum. 
Throughout our work, we express the RCs up to the order of $\Lambda^{-2}$ and consistently neglect higher order corrections.

\subsection{$a_{\mu}$ computation}\label{sec:App_Ba}

We evaluate $a_{\mu}$ from one-loop $\mu\to \mu\gamma$ vertex-function, which requires to consider both $\delta Z_{AA}$ and $\delta Z_{ZA}$ wave-function RCs due to the $Z-\gamma$ mixing.  Their explicit expressions computed from on-shell conditions are given as 

\begin{alignat}{6}
\delta Z_{AA}&=\Big[\frac{g_{_Y}^2g_{_W}^2}{24(g_{_Y}^2+g_{_W}^2)\pi^2}+\frac{5g_{_Y}^2g_{_W}^2B_{0}(M_{_W}^2)}{16(g_{_Y}^2+g_{_W}^2)\pi^2}\nonumber\\
&+\frac{g_{_Y}^2g_{_W}^2B_{1}(M_{_W}^2)}{8(g_{_Y}^2+g_{_W}^2)\pi^2}+\frac{g_{_Y}^2g_{_W}^2M_{_W}^2\,dB_{0}(M_{_W}^2)}{8(g_{_Y}^2+g_{_W}^2)\pi^2}\nonumber\\
&-\frac{g_{_Y}^2g_{_W}^4v^2\,dB_{0}(M_{_W}^2)}{32(g_{_Y}^2+g_{_W}^2)\pi^2}+\frac{g_{_Y}^2g_{_W}^2\,dB_{00}(M_{r^{\pm\pm}}^2)}{(g_{_Y}^2+g_{_W}^2)\pi^2}\nonumber\\
&+\frac{g_{_Y}^2g_{_W}^2\Big(dB_{00}(M_{h^{\pm}}^2)+3\,dB_{00}(M_{_W}^2)\Big)}{4(g_{_Y}^2+g_{_W}^2)\pi^2}\nonumber\\
&+\sum_{l=e,\mu,\tau}\frac{g_{_Y}^2g_{_W}^2B_{1}(M_{_l}^2)}{4(g_{_Y}^2+g_{_W}^2)\pi^2}+\sum_{q_u=u,c,t}\frac{g_{_Y}^2g_{_W}^2B_{1}(M_{q_u}^2)}{3(g_{_Y}^2+g_{_W}^2)\pi^2}\nonumber\\
&+\sum_{q_d=d,s,b}\frac{g_{_Y}^2g_{_W}^2B_{1}(M_{q_d}^2)}{12(g_{_Y}^2+g_{_W}^2)\pi^2}-\sum_{l=e,\mu,\tau}\frac{2g_{_Y}^2g_{_W}^2dB_{00}(M_{_l}^2)}{2(g_{_Y}^2+g_{_W}^2)\pi^2}\nonumber\\
&-\sum_{q_u=u,c,t}\frac{2g_{_Y}^2g_{_W}^2dB_{00}(M_{q_u}^2)}{3(g_{_Y}^2+g_{_W}^2)\pi^2}\nonumber\\
&-\sum_{q_d=d,s,b}\frac{g_{_Y}^2g_{_W}^2dB_{00}(M_{q_d}^2)}{6(g_{_Y}^2+g_{_W}^2)\pi^2}\Big],
\end{alignat}
and 
\begin{alignat}{4}
\delta Z_{AZ} & = \Big[\frac{g_{_Y}g_{_W}^3}{12(g_{_Y}^2+g_{_W}^2)\pi^2}+\frac{g_{_Y}^3g_{_W}\Big(4A_{0}(M_{r^{\pm\pm}}^2)+A_{0}(M_{h^{\pm}}^2)\Big)}{4(g_{_Y}^2+g_{_W}^2)M_{_Z}^2\pi^2}\nonumber\\
&-\sum_{l=e,\mu,\tau}\frac{g_{_Y}g_{_W}\Big(3g_{_Y}^2A_{0}(M_{_l}^2)+g_{_W}^2A_{0}(M_{_l}^2)\Big)}{8(g_{_Y}^2+g_{_W}^2)M_{_Z}^2\pi^2}\nonumber\\
&+\sum_{q_d=d,s,b}\frac{g_{_Y}g_{_W}\Big(3g_{_W}^2A_{0}(M_{q_d}^2)-g_{_Y}^2A_{0}(M_{q_d}^2)\Big)}{24(g_{_Y}^2+g_{_W}^2)M_{_Z}^2\pi^2}\nonumber\\
&+\sum_{q_u=u,c,t} \frac{g_{_Y}g_{_W}\Big(3g_{_W}^2A_{0}(M_{q_u}^2)-5g_{_Y}^2A_{0}(M_{q_u}^2)\Big)}{12(g_{_Y}^2+g_{_W}^2)M_{_Z}^2\pi^2}\nonumber\\
&+\frac{g_{_Y}g_{_W}}{16(g_{_Y}^2+g_{_W}^2)M_{_Z}^2\pi^2}B_{00}(M_{_Z}^2,M_{_W}^2)\Big(10g_{_W}^2M_{_Z}^2\nonumber\\
&+4g_{_W}^2M_{_W}^2+g_{_Y}^2g_{_W}^2v^2-4g_{_Y}^2+10g_{_W}^2\Big)\nonumber\\
&+\frac{g_{_Y}g_{_W}\Big(g_{_Y}^2A_{0}(M_{_W}^2)-5g_{_W}^2A_{0}(M_{_W}^2)\Big)}{8(g_{_Y}^2+g_{_W}^2)M_{_Z}^2\pi^2}\nonumber\\
&+\frac{g_{_Y}^3g_{_W}\Big(B_{00}(M_{_Z}^2,M_{h^{\pm}}^2)-4B_{00}(M_{_Z}^2,M_{r^{\pm\pm}}^2)\Big)}{2(g_{_Y}^2+g_{_W}^2)M_{_Z}^2\pi^2}\nonumber\\
&+\sum_{l=e,\mu,\tau}\frac{g_{_Y}g_{_W}\Big(3g_{_Y}^2-g_{_W}^2\Big)B_{00}(M_{_Z}^2,M_{_l}^2)}{4(g_{_Y}^2+g_{_W}^2)M_{_Z}^2\pi^2}\nonumber\\
&+\sum_{q_d=d,s,b} \frac{g_{_Y}g_{_W}\Big(g_{_Y}^2-3g_{_W}^2\Big)B_{00}(M_{_Z}^2,M_{q_d}^2)}{12(g_{_Y}^2+g_{_W}^2)M_{_Z}^2\pi^2}\nonumber\\
&+\sum_{q_u=u,c,t}\frac{g_{_Y}g_{_W}\Big(5g_{_Y}^2-3g_{_W}^2\Big)B_{00}(M_{_Z}^2,M_{q_u}^2)}{6(g_{_Y}^2+g_{_W}^2)M_{_Z}^2\pi^2}\nonumber\\
&+\sum_{f=l,q_u,q_d}\frac{g_{_Y}g_{_W}\Big(g_{_W}^2-3g_{_Y}^2\Big)B_{1}(M_{_Z}^2,M_{_f}^2)}{8(g_{_Y}^2+g_{_W}^2)M_{_Z}^2\pi^2}\nonumber\\
&+\frac{g_{_Y}g_{_W}^3B_{1}(M_{Z}^2,M_{_W}^2)}{4(g_{_Y}^2+g_{_W}^2)\pi^2}\Big]+\frac{g_{_Y}g_{_W}}{2(g_{_Y}^2+g_{_W}^2)\pi^2}\Big[\nonumber\\
&\Big(\frac{\mathcal{C}_{B\mathcal{R}}-\mathcal{C}_{W\mathcal{R}}}{\Lambda^2}\Big)A_{0}(M_{r^{\pm\pm}}^2)+\Big(\frac{\mathcal{C}_{B\mathcal{S}}-\mathcal{C}_{W\mathcal{S}}}{\Lambda^2}\Big)\nonumber\\
&A_{0}(M_{h^{\pm}}^2)\Big],
\end{alignat}
respectively.

Similarly, we include the wave-function renormalisation for the left and right chiral components of external muon fields 
\begin{alignat}{4}
\delta Z_{f_{_L}} &= \Big[\frac{(g_{_Y}^4+4g_{_Y}^2g_{_W}^2+3g_{_W}^4)}{64(g_{_Y}^2+g_{_W}^2)\pi^2}\nonumber\\
&-\frac{f_{\mathcal{R}}^2 M_{\mu}^2(dB_{0}(M_{\mu}^2,,M_{r^{\pm\pm}}^2)-dB_{1}(M_{\mu}^2,,M_{r^{\pm\pm}}^2))}{4\pi^2}\nonumber\\
&-\frac{f_{\mathcal{S}}^2 \Big(B_{0}(M_{\mu}^2,M_{h^{\pm}}^2)-B_{1}(M_{\mu}^2,M_{h^{\pm}}^2)\Big)}{8\pi^2}\nonumber\\
&-\frac{f_{\mathcal{S}}^2 M_{\mu}^2 \Big(dB_{0}(M_{\mu}^2,M_{h^{\pm}}^2)-dB_{1}(M_{\mu}^2,M_{h^{\pm}}^2)\Big)}{8\pi^2}\nonumber\\
&-\frac{g_{_Y}^2g_{_W}^2}{8(g_{_Y}^2+g_{_W}^2)\pi^2}\Big(B_{0}(M_{\mu}^2,M_{\mu}^2)-B_{1}(M_{\mu}^2,M_{\mu}^2)\nonumber\\
&+2M_{\mu}^2 \Big(dB_{0}(M_{\mu}^2,M_{\mu}^2)-dB_{1}(M_{\mu}^2,M_{\mu}^2)\Big)\Big)\nonumber\\
&-\frac{g_{_W}^2 \Big(g_{_Y}^2+g_{_W}^2 \Big)\Big(B_{0}(M_{\mu}^2,M_{_W}^2)+B_{1}(M_{\mu}^2,M_{_W}^2)\Big)}{16(g_{_Y}^2+g_{_W}^2)\pi^2}\nonumber\\
&-\frac{(g_{_Y}^2-g_{_W}^2 )^2\Big(B_{0}(M_{\mu}^2,M_{_Z}^2)+B_{0}(M_{\mu}^2,M_{_W}^2)\Big)}{32(g_{_Y}^2+g_{_W}^2)\pi^2}\nonumber\\
&-\frac{M_{\mu}^2 \Big(B_{0}(M_{\mu}^2,M_{_H}^2)-B_{1}(M_{\mu}^2,M_{_H}^2)\Big)}{16\pi^2v^2}\nonumber\\&-\frac{M_{\mu}^2 \Big(B_{0}(M_{\mu}^2,M_{_Z}^2)-B_{1}(M_{\mu}^2,M_{_Z}^2)\Big)}{16\pi^2v^2}\nonumber\\
&-\frac{M_{\mu}^4}{8\pi^2v^2}\Big(dB_{0}(M_{\mu}^2,M_{_W}^2)+dB_{1}(M_{\mu}^2,M_{_W}^2)\nonumber\\
&-2dB_{0}(M_{\mu}^2,M_{_H}^2)+dB_{1}(M_{\mu}^2,M_{_H}^2)-dB_{1}(M_{\mu}^2,M_{_Z}^2)\Big)\Big]\nonumber\\
&+\frac{v^2(2f_{\mathcal{R}}^2\mathcal{C}_{\phi\mathcal{R}\mathcal{D}}+f_{\mathcal{S}}^2\mathcal{C}_{\phi\mathcal{S}\mathcal{D}}+f_{\mathcal{S}}\mathcal{C}_{l\phi\mathcal{S}})}{16\pi^2\Lambda^2}\Big[\nonumber\\
&\Big(B_{0}(M_{\mu}^2,M_{r^{\pm\pm}}^2)+B_{1}(M_{\mu}^2,M_{r^{\pm\pm}}^2)\Big)\nonumber\\
&+M_{_\mu}^2\Big(dB_{0}(M_{\mu}^2,M_{r^{\pm\pm}}^2)+dB_{1}(M_{\mu}^2,M_{r^{\pm\pm}}^2)\Big)\Big]\nonumber\\
&+\frac{M_{\mu}^2v^2f_{\mathcal{R}}\mathcal{C}_{e\mathcal{R}\phi}}{4\pi^2\Lambda^2}\Big[\Big(dB_{0}(M_{\mu}^2,M_{r^{\pm\pm}}^2)\nonumber\\
&+dB_{1}(M_{\mu}^2,M_{r^{\pm\pm}}^2)\Big)\Big]+\frac{f_{\mathcal{R}}M_{\mu}^2v^2\mathcal{C}_{l\phi\mathcal{R}}}{4\pi^2\Lambda^2}dB_{0}(M_{\mu}^2,M_{r^{\pm\pm}}^2)\nonumber\\
&+\frac{f_{\mathcal{R}}M_{\mu}v\,\mathcal{C}_{\mathcal{R}le}}{4\sqrt{2}\pi^2\Lambda^2}\Big[B_{0}(M_{\mu}^2,M_{r^{\pm\pm}}^2)+B_{1}(M_{\mu}^2,M_{r^{\pm\pm}}^2)\nonumber\\
&+2\,M_{\mu}^2\,dB_{0}(M_{\mu}^2,M_{r^{\pm\pm}}^2)\Big]+\frac{f_{\mathcal{S}}M_{\mu}v\,\mathcal{C}_{\mathcal{S}le}}{4\sqrt{2}\pi^2\Lambda^2}\nonumber\\
&\Big[B_{0}(M_{\mu}^2,M_{r^{\pm\pm}}^2)+B_{1}(M_{\mu}^2,M_{r^{\pm\pm}}^2)\nonumber\\
&+2\,M_{\mu}^2\,\Big(dB_{0}(M_{\mu}^2,M_{r^{\pm\pm}}^2)+dB_{1}(M_{\mu}^2,M_{r^{\pm\pm}}^2)\Big)\Big],
\end{alignat}
and
\begin{alignat}{4}
\delta Z_{f_{R}} &= \Big[\frac{g_{_Y}^2}{16\pi^2}-\frac{g_{_Y}^4\Big(B_{0}(M_{_Z}^2,M_{\mu}^2)+B_{1}(M_{_Z}^2,M_{\mu}^2)\Big)}{8(g_{_Y}^2+g_{_W}^2)\pi^2}\nonumber\\
&-\frac{f_{\mathcal{R}}^2}{4\pi^2}\Big(B_{0}(M_{r^{\pm\pm}}^2,M_{\mu}^2)+B_{1}(M_{r^{\pm\pm}}^2,M_{\mu}^2)\nonumber\\
&+M_{\mu}^2\Big(dB_{0}(M_{r^{\pm\pm}}^2,M_{\mu}^2),dB_{1}(M_{r^{\pm\pm}}^2,M_{\mu}^2)\Big)\Big)\nonumber\\
&-\frac{f_{\mathcal{S}}^2M_{\mu}^2}{8\pi^2}\Big(dB_{0}(M_{h^{\pm}}^2,M_{\mu}^2)+dB_{1}(M_{h^{\pm}}^2,M_{\mu}^2)\Big)\nonumber\\
&-\frac{g_{_Y}^2g_{_W}^2}{8(g_{_Y}^2+g_{_W}^2)\pi^2}\Big(B_{0}(M_{\mu}^2,M_{\mu}^2)+B_{1}(M_{\mu}^2,M_{\mu}^2)\nonumber\\
&-2\Big(dB_{0}(M_{\mu}^2,M_{\mu}^2)+dB_{1}(M_{\mu}^2,M_{\mu}^2)\Big)\Big)\nonumber\\
&-\frac{M_{\mu}^2}{16\pi^2v^2}\Big(2B_{0}(M_{_W}^2,M_{\mu}^2)+B_{0}(M_{h^{\pm}}^2,M_{\mu}^2)\nonumber\\
&+2B_{1}(M_{_W}^2,M_{\mu}^2)+B_{0}(M_{_Z}^2,M_{\mu}^2)-B_{1}(M_{_Z}^2,M_{\mu}^2)\Big)\nonumber\\
&-\frac{M_{\mu}^2}{32(g_{_Y}^2+g_{_W}^2)\pi^2}\Big(8g_{_Y}^2g_{_W}^2dB_{0}(M_{\mu}^2,M_{\mu}^2)+(3g_{_Y}^4\nonumber\\
&-3g_{_Y}^2g_{_W}^2-g_{_Y}^4)dB_{0}(M_{_Z}^2,M_{\mu}^2)-g_{_Y}^2g_{_W}^2dB_{1}(M_{\mu}^2,M_{\mu}^2)\nonumber\\
&-(5g_{_Y}^4-2g_{_Y}^2g_{_W}^2+g_{_W}^4)dB_{1}(M_{_Z}^2,M_{\mu}^2)\Big)-\frac{M_{\mu}^4}{8\pi^2v^2}\nonumber\\
&\Big(dB_{0}(M_{_W}^2,M_{\mu}^2)+2dB_{0}(M_{h^{\pm}}^2,M_{\mu}^2)+dB_{1}(M_{h^{\pm}}^2,M_{\mu}^2)\nonumber\\
&+dB_{1}(M_{_W}^2,M_{\mu}^2)+dB_{1}(M_{_Z}^2,M_{\mu}^2)\Big)\Big]+\frac{\mathcal{C}_{\phi\mathcal{R}\mathcal{D}}f_{\mathcal{R}}^2v^2}{8\pi^2\Lambda^2}\nonumber\\
&\Big[B_{0}(M_{h^{\pm}}^2,M_{\mu}^2)+B_{1}(M_{h^{\pm}}^2,M_{\mu}^2)+M_{\mu}^2\Big(dB_{0}(M_{h^{\pm}}^2,M_{\mu}^2)\nonumber\\
&dB_{1}(M_{h^{\pm}}^2,M_{\mu}^2)\Big)\Big]+\frac{\mathcal{C}_{\phi\mathcal{S}\mathcal{D}}f_{\mathcal{S}}^2M_{\mu}^2v^2}{16\pi^2\Lambda^2}\Big[dB_{0}(M_{h^{\pm}}^2,M_{\mu}^2)\nonumber\\
&+dB_{1}(M_{h^{\pm}}^2,M_{\mu}^2)\Big]+\frac{\mathcal{C}_{e\mathcal{R}\phi}f_{\mathcal{R}}v^2}{4\pi^2\Lambda^2}\Big[B_{0}(M_{r^{\pm\pm}}^2,M_{\mu}^2)\nonumber\\
&+B_{1}(M_{r^{\pm\pm}}^2,M_{\mu}^2)+M_{\mu}^2\Big(dB_{0}(M_{r^{\pm\pm}}^2,M_{\mu}^2)\nonumber\\
&+dB_{1}(M_{r^{\pm\pm}}^2,M_{\mu}^2)\Big)\Big]+\frac{\mathcal{C}_{l\phi\mathcal{R}}f_{\mathcal{R}}M_{\mu}^2v^2dB_{0}(M_{r^{\pm\pm}}^2,M_{\mu}^2)}{4\pi^2\Lambda^2}\nonumber\\
&+\frac{\mathcal{C}_{l\phi\mathcal{S}}f_{\mathcal{S}}M_{\mu}^2v^2}{16\pi^2\Lambda^2}\Big[dB_{0}(M_{h^{\pm}}^2,M_{\mu}^2)+dB_{1}(M_{h^{\pm}}^2,M_{\mu}^2)\Big]\nonumber\\
&+\frac{f_{\mathcal{R}}M_{\mu}v\,\mathcal{C}_{\mathcal{R}le}}{4\sqrt{2}\pi^2\Lambda^2}\Big[B_{0}(M_{\mu}^2,M_{r^{\pm\pm}}^2)+B_{1}(M_{\mu}^2,M_{r^{\pm\pm}}^2)\nonumber\\
&+2\,M_{\mu}^2\,dB_{0}(M_{\mu}^2,M_{r^{\pm\pm}}^2)\Big]+\frac{f_{\mathcal{S}}M_{\mu}v\,\mathcal{C}_{\mathcal{S}le}}{4\sqrt{2}\pi^2\Lambda^2}\nonumber\\
&\Big[B_{0}(M_{\mu}^2,M_{r^{\pm\pm}}^2)+B_{1}(M_{\mu}^2,M_{r^{\pm\pm}}^2)\nonumber\\
&+2\,M_{\mu}^2\,\Big(dB_{0}(M_{\mu}^2,M_{r^{\pm\pm}}^2)+dB_{1}(M_{\mu}^2,M_{r^{\pm\pm}}^2)\Big)\Big].	
\end{alignat}

The vev-renormalisation term is computed from the Tadpole RC $\delta \text{T}$ as $\delta v = -\delta \text{T} /M_{_H}^2$ (see e.g.~\cite{Denner:2019vbn})
\begin{alignat}{4}
\delta v&=-\frac{1}{M_{_H}^2}\Big[\frac{g_{_W}^2M_{_W}^2v}{16\pi^2}+\frac{(g_{_Y}^2+g_{_W}^2)M_{_Z}^2v}{32\pi^2}\nonumber\\
&+\sum_{l=e,\mu,\tau}\frac{M_{_l}^2A_{0}(M_{_l}^2)}{4\pi^2v}+\sum_{q_u=u,c,t}\frac{3 M_{q_u}^2A_{0}(M_{q_{u}}^2)}{4\pi^2v}\nonumber\\
&+\sum_{q_d=d,s,b}\frac{3 M_{q_d}^2A_{0}(M_{q_{d}}^2)}{4\pi^2v}-\frac{3\lambda_{1}v\,A_{0}(M_{_H}^2)}{16\pi^2}\nonumber\\
&-\frac{v\Big(\lambda_{5}A_{0}(M_{r^{\pm\pm}}^2)+\lambda_{4}A_{0}(M_{h^{\pm}}^2)\Big)}{16\pi^2}-\frac{3g_{_W}^2vA_{0}(M_{_W}^2)}{32\pi^2}\nonumber\\
&-\frac{\lambda_{1}v\Big(2A_{0}(M_{_W}^2)+A_{0}(M_{_Z}^2)\Big)}{16\pi^2}-\frac{3(g_{_Y}^2+g_{_W}^2)vA_{0}(M_{_Z}^2)}{64\pi^2}\nonumber\\
&+\frac{\mathcal{C}_{\phi\mathcal{R}\mathcal{D}}}{\Lambda^2}\Big(\frac{M_{r^{\pm\pm}}v}{16\pi^2}+\frac{\lambda_{5}v^3}{32\pi^2}\Big)A_{0}(M_{r^{\pm\pm}}^2)\nonumber\\
&+\frac{\mathcal{C}_{\phi\mathcal{S}\mathcal{D}}}{\Lambda^2}\Big(\frac{M_{h^{\pm}}v}{16\pi^2}+\frac{\lambda_{4}v^3}{32\pi^2}\Big)A_{0}(M_{h^{\pm}}^2)\nonumber\\
&+\frac{\mathcal{C}_{\phi\mathcal{R}}}{\Lambda^2}\frac{v^3A_{0}(M_{r^{\pm\pm}}^2)}{16\pi^2}+\frac{\mathcal{C}_{\phi\mathcal{S}}}{\Lambda^2}\frac{v^3A_{0}(M_{h^{\pm}}^2)}{16\pi^2}\Big].	
\end{alignat}

The divergences associated with the Wilson-coefficients are removed through $\overline{\text{MS}}$ renormalisation of $\mathcal{C}_{eA}$
\begin{alignat}{4}
\delta \mathcal{C}_{eA} &= \Big[\frac{f_{\mathcal{S}}^2}{32\pi^2}+\frac{f_{\mathcal{R}}^2}{16\pi^2}+\frac{g_{_W}^2}{64\pi^2}+\frac{5g_{_Y}^4}{128(g_{_Y}^2+g_{_W}^2)\pi^2}\nonumber\\
&+\frac{65g_{_Y}^2g_{_W}^2}{192(g_{_Y}^2+g_{_W}^2)\pi^2}+\frac{g_{_W}^4}{128(g_{_Y}^2+g_{_W}^2)\pi^2}-\frac{3\lambda_{1}}{16\pi^2}\nonumber\\
&-\frac{\lambda_{1}M_{_W}^2}{8M_{_H}^2\pi^2}-\frac{\lambda_{1}M_{_Z}^2}{16M_{_H}^2\pi^2}-\frac{(\lambda_{5}M_{r^{\pm\pm}}^2+\lambda_{4}M_{h^{\pm}}^2)}{16M_{_H}^2\pi^2}\nonumber\\
&-\frac{3g_{_W}^2M_{_W}^2}{32M_{_H}^2\pi^2}-\frac{3(g_{_Y}^2+g_{_W}^2)M_{_Z}^2}{64M_{_H}^2\pi^2}-\sum_{l=e,\mu,\tau}\frac{M_{_l}^4}{4M_{_H}^2\pi^2v^2}\nonumber\\
&-\sum_{q_u=u,c,t}\frac{M_{q_u}^4}{4M_{_H}^2\pi^2v^2}-\sum_{q_d=d,s,b}\frac{M_{q_d}^4}{4M_{_H}^2\pi^2v^2}\Big]\frac{\mathcal{C}_{eA}}{\Lambda^2}\nonumber\\
&+\Big[\frac{g_{_Y}g_{_W}^3M_{_W}^2}{8(g_{_Y}^2+g_{_W}^2)M_{_Z}^2\pi^2}+\frac{g_{_Y}^3g_{_W}^3v^2}{32(g_{_Y}^2+g_{_W}^2)M_{_Z}^2\pi^2}\Big]\frac{\mathcal{C}_{eZ}}{\Lambda^2}\nonumber\\
&+\frac{f_{\mathcal{S}}M_{\mu}}{32\sqrt{g_{_Y}^2+g_{_W}^2}\pi^2v}\Big(g_{_Y}\frac{\mathcal{C}_{eW\mathcal{S}}}{\Lambda^2}+2g_{_W}\frac{\mathcal{C}_{eB\mathcal{S}}}{\Lambda^2}\Big),
\end{alignat}

\subsection{Higgs decay}\label{sec:App_Bb}

In this section we have listed down the renormalisation constants (RC) computed for the case of 125 GeV Higgs decay to di-photon mentioned in Sec.~\ref{sec:hdec}. Here we only mention the part of the RCs which arise due to the divergence coming through the coupling of $r^{\pm\pm}$ to other fields. The parts of the RCs containing the singly charged field $h^{\pm}$ have been documented in the Appendix of Ref.~\cite{Anisha:2021fzf}.
The on-shell wave function RCs for the external fields are given as    
\begin{alignat}{4}
\delta Z_{AA,r^{\pm\pm}}&=\frac{g_{_Y}^2\,g_{_W}^2\,dB_{00}(M_{r^{\+\+}}^2)}{ \pi ^2 \left(g_{_Y}^2+g_{_W}^2\right)}\nonumber\\
&-\frac{
	A_{0}(M_{r^{\+\+}}^{2}) \left(\mathcal{C}_{B\mathcal{R}} \,g_{_W}^2+\mathcal{C}_{W\mathcal{R}} \,g_{_Y}^2\right)}{4 \pi ^2 \left(g_{_Y}^2+g_{_W}^2\right)\Lambda^2},		
\end{alignat}

\begin{alignat}{5}
\delta Z_{ZA,r^{\pm\pm}}&=\frac{g_{_Y} g_{_W}\Big(
	g_{_Y}^2 B_{00}(M_{r^{\pm\pm}}^2)- g_{_Y}^2 A_{0}(M_{r^{\pm\pm}}^2)\Big)}{\pi ^2 M_{_Z}^2 (g_{_Y}^2+g_{_W}^2)\Lambda^2},
\end{alignat}
and	
\begin{alignat}{5}
\delta Z_{H,r^{\pm\pm}}&=	\frac{1}{64 \pi ^2\Lambda^2}\Big(4\mathcal{C}_{\phi\mathcal{R}\mathcal{D}} \,\lambda_{5}^2 \,v^4 \,dB_{0}(M_{H}^2,M_{r^{\pm\pm}}^2) \nonumber\\
&+8 \,\mathcal{C}_{\phi\mathcal{R}}
\,\lambda_{5}\,v^4 \,dB_{0}(M_{H}^2,M_{r^{\pm\pm}}^2)\nonumber\\
&+8 \,v^2 \,\mathcal{C}_{\phi\mathcal{R}\mathcal{D}} \,\lambda_{5}
B_{1}(M_{H}^2,M_{r^{\pm\pm}}^2)\nonumber\\
&+8 \,\mathcal{C}_{\phi\mathcal{R}\mathcal{D}} \,\lambda_{5} \,M_{r^{\pm\pm}}^2 \,v^2
\,dB_{0}(M_{H}^2,M_{r^{\pm\pm}}^2)\nonumber\\
&+8 \,\mathcal{C}_{\phi\mathcal{R}\mathcal{D}} \,\lambda_{5} M_{H}^2 \,v^2
\,dB_{1}(M_{H}^2,M_{r^{\pm\pm}}^2)\nonumber\\
&+4 \,\mathcal{C}_{\mathcal{R}\phi\mathcal{D}}
A_{0}(M_{r^{\pm\pm}}^2)\Big).
\end{alignat}	
The tadpole counter term is also computed in a similar fashion as given in Sec.~\ref{sec:App_Ba}
\begin{alignat}{6}
\delta v_{r^{\pm\pm}} &=-\frac{A_{0}(M_{r^{\pm\pm}}^2)}{64 \pi ^2 M_{H}^2 \Lambda^2 }  \Big(4 \,v^3 \,\mathcal{C}_{\phi\mathcal{R}} +2\lambda_{5} \,v^2 \,\mathcal{C}_{\phi\mathcal{R}\mathcal{D}} \nonumber\\
&+4 \,M_{r^{\pm\pm}}^2 \,\mathcal{C}_{\phi\mathcal{R}\mathcal{D}}
-4 \,v \lambda_{5}\,\Lambda^2\Big).
\end{alignat}
The RCs associated with the Wilson coefficients are given by  	
\begin{alignat}{6}
\delta\mathcal{C}_{A\phi,r^{\pm\pm}} & = \frac{\lambda_{5}\,(g_{_Y}^2\,\mathcal{C}_{W\mathcal{R}}+g_{_W}^2\,\mathcal{C}_{B\mathcal{R}})}{16(g_{_Y}^2+g_{_W}^2)\pi^2\Lambda^2},
\end{alignat}
and	
\begin{alignat}{7}
\delta\mathcal{C}_{\widetilde{A}\phi,r^{\pm\pm}} & = \frac{\lambda_{5}\,(g_{_Y}^2\,\mathcal{C}_{\widetilde{W}\mathcal{R}}+g_{_W}^2\,\mathcal{C}_{\widetilde{B}\mathcal{R}})}{16(g_{_Y}^2+g_{_W}^2)\pi^2\Lambda^2}.
\end{alignat}

\bibliography{paper.bbl} 
	
\end{document}